\documentclass[final]{jfm}
\usepackage{graphicx}

\ifCUPmtlplainloaded \else
  \checkfont{eurm10}
  \iffontfound
    \IfFileExists{upmath.sty}
      {\typeout{^^JFound AMS Euler Roman fonts on the system,
                   using the 'upmath' package.^^J}%
       \usepackage{upmath}}
      {\typeout{^^JFound AMS Euler Roman fonts on the system, but you
                   dont seem to have the}%
       \typeout{'upmath' package installed. JFM.cls can take advantage
                 of these fonts,^^Jif you use 'upmath' package.^^J}%
       \providecommand\upi{\pi}%
      }
  \else
    \providecommand\upi{\pi}%
  \fi
\fi

\ifCUPmtlplainloaded \else
  \checkfont{msam10}
  \iffontfound
    \IfFileExists{amssymb.sty}
      {\typeout{^^JFound AMS Symbol fonts on the system, using the
                'amssymb' package.^^J}%
       \usepackage{amssymb}%
         \let\leq=\leqslant
         
      }{}
  \fi
\fi

\ifCUPmtlplainloaded \else
  \IfFileExists{amsbsy.sty}
    {\typeout{^^JFound the 'amsbsy' package on the system, using it.^^J}%
     \usepackage{amsbsy}}
    {\providecommand\boldsymbol[1]{\mbox{\boldmath $##1$}}}
\fi

\providecommand\bnabla{\boldsymbol{\nabla}}
\providecommand\bcdot{\boldsymbol{\cdot}}

\newcommand\dr{\mathrm{d}}
\newcommand\er{\mathrm{e}}
\newcommand\ir{\mathrm{i}}

\newcommand\Ub{\boldsymbol{U}}
\newcommand\xb{{\boldsymbol x}}
\newcommand\kb{{\boldsymbol k}}

\newcommand\kpb{\boldsymbol{k^{\prime}}}
\newcommand\lb{\boldsymbol{l}}

\def\sfbsQ{\mathsfbi{Q}}
\def\sfbsS{\mathsfbi{S}}
\def\sfbsI{\mathsfbi{I}}

\newcommand\Real{\mbox{Re}} % cf plain TeX's \Re and Reynolds number
\newcommand\Imag{\mbox{Im}} % cf plain TeX's \Im
\newcommand\Frou{\mbox{\textit{Fr}}}  % Reynolds number
\newsavebox{\astrutbox}
\sbox{\astrutbox}{\rule[-5pt]{0pt}{20pt}}

\newcommand\etal{\mbox{\textit{et al.}}}

\title[Current effects on scattering of surface gravity waves by bottom topography]%
{Current effects on scattering of surface gravity waves by bottom topography}

\author[R. Magne and F. Ardhuin]%
{R\ls U\ls D\ls Y\ns M\ls A\ls G\ls N\ls E$^{1,2}$\ns \and F\ls A\ls B\ls R\ls
I\ls C\ls E\ns A\ls R\ls D\ls H\ls U\ls I\ls N$^1$}
 \affiliation{$^1$Centre Militaire
d'Oc{\'e}anographie, Service Hydrographique et Oc{\'e}anographique de la
Marine, 29275 Brest, France
 \\[\affilskip]
 $^2$Laboratoire de Sondages Electromagn{\'e}tique de l'Environnement Terrestre,
Universit{\'e} de Toulon et du Var, La Garde, France
\\[\affilskip]
ardhuin@shom.fr}

\date{8 September 2005 (draft)}%% and in revised form 23 June 2001}
\begin{document}

\maketitle

%%% Abstract for HAL
%Scattering of random surface gravity waves by small amplitude topography in the
%presence of a  uniform current is investigated theoretically. This problem is
%relevant to ocean waves propagation on shallow   continental shelves where
%tidal   currents are often significant. A perturbation expansion of the wave
%action to second order in powers of the bottom amplitude yields an evolution
%equation for the wave action spectrum.  A scattering source term gives the rate
%of exchange of the wave action spectrum between wave components, with
%conservation of the total action at each absolute frequency. With and without
%current, the scattering term yields reflection coefficients for the amplitudes
%of waves that converge, in the limit of small bottom amplitudes and small
%Froude numbers, to the results of previous theories for monochromatic waves
%propagating in one dimension over sinusoidal bars. Over sandy continental
%shelves, tidal currents are known to generate sandwaves with scales comparable
%to those of surface waves.   Application of the theory to such a real
%topography   suggests that scattering mainly results in a broadening of the
%directional wave spectrum, due to forward scattering, while the back-scattering
%is generally weaker. The current may strongly influence surface gravity wave
%scattering by selecting different bottom scales with widely different spectral
%densities due the sharp bottom spectrum roll-off.

\begin{abstract}
  The scattering of random surface gravity waves by topography of small amplitude, and horizontal scales of the order of the
  wavelength,  is investigated theoretically in the presence of a
  an almost uniform irrotational current. This problem is relevant to ocean waves propagation on shallow
  continental shelves where tidal
  currents are often significant. Defining the small scale bottom
  amplitude normalized by the mean water depth, $\eta=h/H$, a  perturbation expansion of the wave action to order $\eta^2$
  yields an evolution equation for the wave action spectrum. Based on numerical calculations for
  sinusoidal bars, a mixed surface-bottom bispectrum, that arises at order $\eta$, is unlikely
   to be significant in most
  oceanic conditions. Neglecting that term, the present theory yields a closed equation with a scattering source term
  that gives the rate of exchange of action between spectral wave components that have the same absolute frequency.
  This source term is proportional to the bottom elevation variance at the resonant wavenumbers, and thus represents a
  Bragg scattering approximation. With current, the source term
  formally combines a direct effect of the bottom
  topography with an indirect effect of the bottom through the modulation of
  the surface current and mean surface elevation.
  For Froude numbers of the order of 0.6 or less, the bottom topography effects dominate. For all Froude
  numbers,  the reflection coefficients for the wave amplitudes that are inferred from the source term are
   asymptotically identical, as $\eta$ goes to zero, to previous theoretical results for monochromatic waves
   propagating in one dimension over sinusoidal bars. In particular, the frequency of the waves that experience the maximum reflection
  is shifted by the current, as the surface wavenumber $k$ changes for a fixed absolute frequency.
  Over sandy continental shelves, tidal currents are known to generate sandwaves with scales comparable to
  those of surface waves, with bottom elevation spectra that roll-off sharply at high wavenumbers.
  Application of the theory to such a real topography suggests that scattering mainly results in a broadening of the
  directional wave spectrum, i.e. forward scattering, while back-scattering is generally weaker. The current may strongly influence surface
   gravity wave scattering by selecting different bottom scales, with widely different spectral densities due the
  sharp bottom spectrum roll-off.
\end{abstract}

\section{Introduction}
%\#\# Ajouter: commentaire sur Miles 1998, Howe 1971
%\#\# Au lieu de Mei 1985: finite detuning, Liu 1987 is wrong, voir aussi Kirby
%1993, Ruban 2004
 Following the early observations of Heathershaw
(1982\nocite{Heathershaw1982}), a considerable body of knowledge has been
accumulated on the scattering of small amplitude surface gravity waves by
periodic bottom topography. An asymptotic theory for small bottom amplitudes,
that reproduces the observed scattering of monochromatic waves over a few
sinusoidal bars, was put forward by Mei (1985\nocite{Mei1985}), leading to
practical phase-resolving equations that may be used to model this phenomenon
for more general bottom shapes (Kirby 1986\nocite{Kirby1986a}). For sinusoidal
bottoms of wavenumber $l$, Mei (1985) proposed an approximate analytical
solution. In two dimensions (one horizontal and the vertical) this solution
yields simple expressions for the wave amplitude reflection coefficient $R$, as
a function of the mismatch between the wavenumber of the surface waves $k$ and
the resonant value $l/2$, for which $R$ is maximum due to Bragg resonance.
Beyond a cut-off value of that mismatch, it was found that the incident and
reflected wave amplitudes oscillate in space instead of decreasing
monotonically from the incident region. In three dimensions the Bragg resonance
condition becomes $\kb = \lb + \kb'$ and $\omega=\omega'$, with $\omega$ and
$\omega'$ the wave radian frequencies corresponding to the wavenumber vectors
$\kb$ and $\kpb$ through the linear dispersion relation.

Other contributions have shown that higher-order theories are necessary to
represent the sub-harmonic resonance observed over a bottom that is a
superposition of two components of different wavelengths (Guazzelli, Rey \&
Belzons 1992\nocite{Guazzelli&al.1992}). Such sub-harmonic resonance was found
to have as large an effect as the lowest order resonance for bottom amplitudes
of only 25\% of the water depth, due to a general stronger reflection for
relatively longer waves. However, these amplitude evolution equations are still
prohibitively expensive for investigating the propagation of random waves over
distances larger than about 100 wavelengths, and the details of the bottom are
typically not available over large areas. Besides, a consistent phase-averaged
wave action evolution equation is also necessary for the investigation of the
long waves associated with short wave groups (Hara \& Mei
1987\nocite{Hara&Mei1987}).

The large scale behaviour of the wave field may rather be represented by the
evolution of the wave action spectrum assuming random phases. Such an approach
was already proposed by Hasselmann (1966\nocite{Hasselmann1966}) and Elter \&
Molyneux (1972\nocite{Elter&Molyneux1972}) for the calculation of wind-wave and
tsunami propagation. A proper theory for the evolution of the wave spectrum can
be obtained from a solvability condition, a method similar to that of Mei
(1985) and Kirby (1988), but applied to the action spectral densities instead
of the amplitudes of monochromatic waves. In the absence of currents the
correct form of that equation was first obtained by Ardhuin \& Herbers
(2002\nocite{Ardhuin&Herbers2002}, hereinafter referred to as AH) using a two
scale approach. They decomposed the water depth $H-h$ in a slowly varying depth
$H$, that causes shoaling and refraction, and a rapidly varying perturbation
$h$ with zero mean, that causes scattering. This equation is formally similar
to general transport equations for waves in random media (e.g. Ryzhik,
Papanicolaou \& Keller 1996\nocite{Ryzhik&al.1996}), although the waves
considered here propagate only in the two horizontal dimensions. The resulting
scattering was shown to be consistent with the dramatic increase of the
directional width of the wave spectra observed on the North Carolina
continental shelf
 (Ardhuin \etal~2003a, 2003b\nocite{Herbers&al.2000a, Ardhuin&Herbers2002,Ardhuin&al.2003a,Ardhuin&al.2003b}).

Recently,  Magne \etal~(2005, hereinafter referred to as
MAHR)\nocite{Magne&al.2005} showed that AH's theory gives the same damping of
incident waves as the Green function solution of Pihl, Mei \& Hancock (2002),
applied to any two dimensional topography, random or not. Investigating the
applicability limits of the scattering term of AH, MAHR also performed
numerical calculations, comparing AH's theory to the accurate matched-boundary
model of Rey (1992\nocite{Rey1992}) that uses a decomposition of the bottom in
a series of steps, including evanescent modes. The numerical results show that
AH's theory is generally limited by the relative bottom amplitude
$\eta=\max(h)/H$ rather than the bottom slope. In particular, AH's theory
predicts accurate reflections, with a relative error of order $\eta$, even for
isolated steps that have an infinite slope (MAHR).

The resulting expression of the Bragg scattering term is consistent with
results for scattering of acoustic and electromagnetic waves obtained by the
small perturbation method, valid in the limit of small $k \max(h)$ where $k$ is
the wavenumber of the propagating waves (Rayleigh 1896\nocite{Rayleigh1896},
see Elfouhaily \& Guerin 2004\nocite{Elfouhaily&Guerin2004} for a review of
this and other approximations). Since there is no scattering for $kH \gg 1$, as
the waves do not `feel' the bottom, the small parameter $\eta=\max(h)/H$ may be
used in the context of surface gravity waves, instead of the more general $k
\max(h)$. For $\eta \ll 1$, the scattering strength is thus entirely determined
by the bottom elevation variance spectrum at the bottom scales resonant with
the incident waves.

Based on these results, Mei's (1985) theory should yield the same reflection
coefficient as AH's theory in the limit of small bottom amplitudes. Yet, AH
predict that the wave amplitude in 2D would decay monotonically, which is not
compatible with the oscillatory nature of Mei's theory for large detunings from
resonance. Further, outside of the surf zone and the associated multiple bar
systems, the application of AH's theory is most relevant in areas where the
bottom topography changes significantly on the scale of the wavelengths of
swells. This often corresponds, over sand, to the presence of sandwaves. These
sandwaves are generated by currents, and particularly by tidal currents (e.g.
Dalrymple Knight \& Lambiase 1978; Idier, Erhold \& Garlan
2002\nocite{Dalrymple&al.1978,Idier&al.2002}). It is thus logical to include
the effects of currents in any theory for wave scattering over a random bottom.
Kirby (1988\nocite{Kirby1988}) developed such a theory for monochromatic waves
over a sinusoidal bottom and a slowly varying mean current, extending Mei's
(1985) work. The geometry of the resonant wavenumbers is modified in that case,
with with incident and reflected waves having the same absolute frequency, but
different wavenumber magnitudes if incident and reflected waves propagate at
different angles relative to the current direction.
 Kirby (1988) also considered the short scale fluctuations of the current, due to the sinusoidal bottom, that may be interpreted as a separate scattering mechanism, and generalized
 further to any irrotational current fluctuations, leading to results similar
 to those obtained for gravity-capillary waves by Bal \& Chou (2002\nocite{Bal&Chou2002}).

 The present paper thus deals with these two questions. An extension of AH's theory for surface gravity wave
 scattering in the presence of irrotational currents with uniform mean velocities is provided in ~\S~2,
 and the differences between this theory and those of Mei
 (1985) and Kirby (1988) are discussed in detail in ~\S~3. Expected oceanographic effects of
 scattering in the presence of a  current are investigated in ~\S~4,
 using a spectral phase-averaged numerical model, predicting the evolution of the wave
 action spectrum, and detailed measurements of the topography in the southern
 North Sea. Conclusions follow in ~\S~5.

\section{Theory}
\subsection{General formulation}
The variation in the action spectral density due to wave-bottom scattering is
derived following the method of AH, now including the effect of a uniform mean
current. The method is identical to that of Kirby (1988) with the difference
that an equation for the spectral wave action is sought instead of one for the
wave amplitudes. Thus intermediate results are identical to those of Kirby
(1988). Since the wave action is a quadratic function in the wave amplitude, we
will naturally consider the wave potential up to second order in the normalized
bottom amplitude $\eta$, in order to have all wave action terms to order
$\eta^2$. The only important terms in this type of calculation are the `secular
terms', i.e. the harmonic oscillator solutions for the wave potential forced at
resonance, with an amplitude that grows unbounded in time. We shall thus obtain
a rate of change of the action from the equality of all the secular terms. The
particularity of the random wave approach is also that we will consider all
possible couplings between wave components, and not just two wave trains. With
random waves, secularity is limited to a sub-space of the wavenumber plane that
generally has a zero measure. Thus the near-resonant terms, once integrated
across the resonant singularity, are the ones that provide the secular terms
for random waves. This integration assumes that the spectral properties are
continuous, a real theoretical problem for nonlinear wave-wave interactions
(e.g. Benney \& Saffmann 1966\nocite{Benney&Saffman1966,Onorato&al.2004},
Onorato \etal~2004). Here we shall see that the only relevant condition is that
the bottom spectrum be continuous, at least in one dimension. This is obviously
satisfied by any real topography, since a truly infinite sinusoidal bottom of
wavelength $L$, with an infinite spectral density at the wavenumber $2 \upi /
L$, is not to be found, even in the laboratory.

 We consider weakly nonlinear random waves propagating over an
irregular bottom with a constant mean depth $H$ and mean current ${\mathbf U}$,
and random topography $h(\xb)$, with $\xb$ the horizontal position vector, so
that the bottom elevation is given by $z=-H+h(\xb)$ where $z$ is the elevation
relative to the still water level. The bottom undulations cause a stationary
random small-scale current fluctuation $({\mathbf u}(\xb,z),w(\xb,z))$ deriving
from a potential $\phi_c$. The free surface is at
$z=\zeta(\xb,t)=\zeta(\xb,t)$. Extension to mean current and mean depth
variations on a large scale follows from a standard two-scale approximation,
identical to that of by Kirby (1988). This is not included in the present
derivation for the sake of simplicity.

The maximum surface slope is characterized by $\varepsilon$ and we shall assume
that $\varepsilon^3 \ll \eta^2$ so that the bottom scattering contributions to
the wave action  to order $\eta^2$ are much larger than the resonant non-linear
four wave interactions (Hasselmann 1962\nocite{Hasselmann1962}) that shall be
neglected. Such interactions could also be allowed in the present calculation
providing an additional source of scattering with the known form due to cubic
non-linearities. For shallow water waves ($kH << 1$) a stricter inequality is
needed to prevent triad wave-wave interactions to enter the action evolution
equation at the same order as bottom scattering.
%%%%%%%%%%%%%%%%%%%%%%%%%%%%%%%%%%%%%%%%%%%%%%%%%%%%%%%%%%%%%%%%%%%%%%%%%%%%%%%%
\begin{figure}
%\epsfig{file=bottomdef.eps,width=\linewidth}
%\centerline{\includegraphics[width=\textwidth]{bottomdef.eps}}
\centerline{\includegraphics[width=\textwidth]{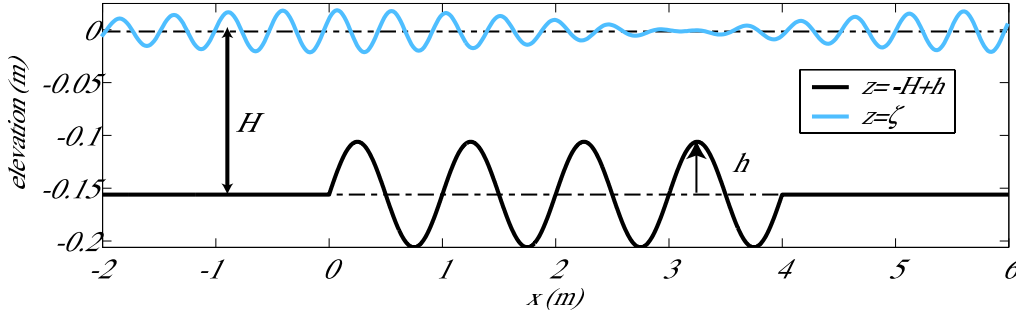}}
\caption{Definition sketch of the mean water depth $H$, and relative bottom
elevation $h$, for one particular case of a sinusoidal bottom investigated in
~\S~3.} \label{hetH}
\end{figure}
%%%%%%%%%%%%%%%%%%%%%%%%%%%%%%%%%%%%%%%%%%%%%%%%%%%%%%%%%%%%%%%%%%%%%%%%%%%%%%%%

The solution is obtained in a frame of reference moving with the mean current
vector $\Ub$, which has the advantage of removing the convective terms due to
the mean current velocity. The corresponding transformation of the horizontal
coordinates is $\xb'=\xb+\Ub t$, where $\xb$ and $\xb'$ are the coordinates in
the moving and fixed frames, respectively. As a result of this transformation,
the bottom is moving, and the bottom boundary condition for the velocity
potential is modified. The governing equations consist of Laplace's equation
for the velocity potential, which includes both wave and current motions, the
bottom kinematic boundary conditions, and Bernoulli's equation with the free
surface kinematic boundary condition. Assuming that the atmospheric pressure is
zero for simplicity, and neglecting surface tension, one has
\begin{eqnarray}
\bnabla^{2} \phi +\frac{\partial ^{2}\phi}{\partial z^{2}} & = & 0\quad
\mbox{for} \quad -H+h \leq z \leq \zeta, \label{laplace} \\
\frac{\partial \phi}{\partial z} &= &\frac{\partial h}{\partial t}+\bnabla \phi
\cdot \bnabla h \quad \mbox{at} \quad z=-H+h,
\label{fond}\\
\frac{\partial{\phi}}{\partial{t}}+g\zeta& = & -\frac{1}{2} \left|\bnabla \phi
\right|^2  - \frac{1}{2} \left(\frac{\partial\phi}{\partial z}\right)^2 +c (t)
\quad \mbox{at} \quad z=\zeta. \label{Bernoulli} \\
\frac{\partial\phi}{\partial z}&=& \frac{\partial{\zeta}}{\partial{t}}+
\bnabla\phi \cdot \bnabla\zeta \quad \mbox{at} \quad  z=\zeta, \label{surf_cin}
\end{eqnarray}
with $c(t)$ a function of time only, to be determined. The symbol $\bnabla$
represents the usual gradient operator restricted
 to the two horizontal dimensions.
 The latter two equations
may be combined to remove the linear part in $\zeta$. Taking $\partial
(\ref{Bernoulli})/\partial t $ +$g$(\ref{surf_cin}), yields,
\begin{eqnarray}
\frac{\partial^2{\phi}}{\partial{t^2}}+g\frac{\partial\phi}{\partial z}=
g\bnabla\phi \cdot \bnabla\zeta - \frac{\partial \zeta}{\partial
t}\frac{\partial^2 \phi}{\partial z \partial t}- \left(1+\frac{\partial
\zeta}{\partial t}\frac{\partial }{\partial z}\right) \left[\bnabla\phi \cdot
\frac{\partial{\bnabla\phi}}{\partial t} + \frac{\partial\phi}{\partial
z}\frac{\partial^2\phi}{\partial t\partial z}\right]&+&c'(t), \nonumber \\
\quad \mbox{at} \quad  z&=&\zeta. \label{comb}
\end{eqnarray}

Following Hasselmann (1962)\nocite{Hasselmann1962}, we approximate $h$ and
$\phi$ with discrete sums over Fourier components, and take the limit to
continuous integrals after deriving expressions for the evolution of the
phase-averaged wave action. We look for a velocity potential solution in the
form
\begin{equation}\label{phi}
  \phi(\xb,z,t)=\sum_{\kb,s} \widehat{\Phi}^s_{\kb}(z,\gamma t)\er^{\ir [\kb \bcdot \xb -s\sigma t]}
  = \sum_{\kb,s} \Phi^s_{\kb}(t)\frac{\cosh\left[k(z+H)\right]}{\cosh(kH)}\er^{\ir \kb \bcdot
  \xb}+\ldots,\label{Spectralplus}
  \end{equation}
where $\sigma$ is the radian frequencies in the moving frame, $\kb$ is the
surface wavenumber, with magnitude $k$, and $s$ is a sign index equal to 1 or
$-1$. In the moving frame of reference, $s=1$ for wave components that
propagate in the direction of the vector $\kb$, and $s=-1$ for components that
propagate in the opposite direction. Thus the radian frequency in the fixed
frame is $\omega=\sigma+s \kb \bcdot \Ub$. The amplitudes
$\widehat{\Phi}^s_{\kb}$ are slowly modulated in time, with a slowness defined
by the small parameter $\gamma$. Because $\phi$ is a real quantity we also have
$\overline{\widehat{\Phi}^s_\kb}=\widehat{\Phi}^{-s}_{-\kb}$, where the overbar
denotes the complex conjugate. Thus the double decomposition made in
(\ref{Spectralplus}) in wavenumber ${\kb}$ and propagation direction $+$ or $-$
replaces a more general decomposition in wavenumber and frequency that would be
necessary if nonlinear dispersive effects were included. Here the frequency
$\sigma$ is always related to $k$ via the linear dispersion relation.

 In the alternative decomposition with
amplitudes $\Phi^s_{\kb}$ that contain the fast time variation, only the part
of the solution that has the vertical structure of Airy waves has been given
explicitly. The other part, represented by `$\ldots$', will be found to be
negligible for small bottom amplitudes. Our rather archaic use of the $s$ index
to distinguish the wave propagation direction is preferred here to the more
modern use of the Hamiltonian variables that combine elevation and potential at
the free surface, widely used for wave-wave interaction studies (e.g. Janssen
2004\nocite{Janssen2004}). The complexity of the Hamiltonian variables appears
unnecessary for the linear waves considered here.

Expanding the bottom boundary condition and wave potential in powers of
$\eta=\max(h)/H$,
\begin{equation}
\phi=\phi_0+ \phi_1+\phi_2+ \ldots, \label{Taylor}
\end{equation}
where each term $\phi_i$ is of order $\eta^i$. The boundary conditions
(\ref{comb}) and (\ref{fond}) are expressed at $z=0$ and $z=-H$, respectively,
using Taylor series of $\phi$ about $z=-H$ and $z=0$.

Unless stated otherwise, these potential amplitudes will be random variables.
Since we are solving for $\phi$ seeking an equation the for wave action $N$, we
must relate $N$ to $\phi$. Accurate to second order in $\varepsilon$ and $\eta$
(see Andrews \& McIntyre 1978 for the general expression of $N$) we have
$N=E/\sigma$ for a monochromatic wave of surface elevation variance $E$ and
intrinsic frequency $\sigma$, in which, following the common usage in
non-accelerated reference frames, the gravity $g$ is left out, so that the
action has units of meters squared times second. For general waves, the
variance $E$ may be written as
\begin{eqnarray}
E(t)&=&\left<\left(\zeta_0+ \zeta_1+\zeta_2+ \ldots \right)^2\right>
=\left<\zeta^2_0+ 2 \zeta_0  \zeta_1 + \left(\zeta^2_1 + 2 \zeta_0
\zeta_2\right) + \ldots \right>, \label{Eexp}
\end{eqnarray}
where $\langle \cdot \rangle$ denotes an average over flow realizations, and
$\zeta_i$ is the surface elevation solution of order $\eta^i$, and terms of
like order in $\eta$ have been grouped. Each of these terms may be expanded in
this form
\begin{equation}
\left<\zeta^2_0\right>=\sum_{\kb,s} \left|Z^s_{0,\kb}\right|^2  = 2 \sum_{\kb}
Z^+_{0,\kb} Z^-_{0,-\kb}
\end{equation}
 For free wave components, the elevation amplitude is proportional to the velocity potential amplitude
 \begin{equation}
Z^s_{i,\kb}=\ir s \sigma \Phi_{j,\kb}^{s}/g \label{pot_to_elev}
\end{equation}
  so that the elevation
co-variances are proportional to the
 co-variances $F^{\Phi}_{i,j,k}$
 of the surface velocity potential,
\begin{equation}\label{Fijkphi}
F^{\Phi}_{i,j,k}=\langle \Phi_{i,\kb}^{+}\Phi_{j,-\kb}^{-}
+\Phi_{i,-\kb}^{-}\Phi_{j,\kb}^{+} \rangle,\label{covarPhi}
\end{equation}
 The
contribution of the complex conjugate pairs of components ($\kb,+$) and
($-\kb,-$) are combined in (\ref{Fijkphi}) so that the covariance
$F^{\Phi}_{i,j,k}$ correspond to that of all waves with wavenumber magnitude
$k$ propagating in the direction of $\kb$. In the limit of small wavenumber
separation, a continuous slowly-varying cross-spectrum can be defined (e.g.
Priestley 1981, ch.11\nocite{Priestley1981}; see also AH),
\begin{equation}\label{cov2}
F^\Phi_{i,j}(\kb)=\lim _{|\Delta k| \rightarrow
0}\frac{F^{\Phi}_{i,j,k}}{\Delta k_x \Delta k_y}.
\end{equation}
The definition of all spectral densities are chosen so that the integral over
the entire wavenumber plane yields the total covariance of $\phi_i$ and
$\phi_j$.

 Finally, $N_{i,j}(\kb)$ is defined as the $(i+j)^{\mathrm{th}}$ order depth-integrated wave
 action contribution from correlation between $i^{\mathrm{th}}$ and $j^{\mathrm{th}}$
order components with wavenumber $k$. From (\ref{Eexp}) and (\ref{pot_to_elev})
one has,
\begin{equation}\label{nrj_ij}
  N_{i,j}(\kb)=\frac{k}{g \sigma} F^\Phi_{i,j}(\kb)\tanh(kH).\label{NfromPhi}
\end{equation}
The spectral wave action is thus,
\begin{equation}\label{nrj}
  N(\kb)=\sum_{i=0}^{\infty} N_i (\kb) =  \sum_{i=0}^{\infty}  \sum_{j=0}^i
  N_{i,i-j}\label{Action_exp}
  (\kb).
\end{equation}

Defining $G_{\lb}$ as the amplitude of the Fourier component of wavenumber
$\lb$, the bottom elevation is given by
\begin{equation}
h(\xb)=\sum_{\lb} G_{\lb} \er^{\ir \lb \bcdot [\xb+\Ub t]},
\end{equation}
with a summation on the entire wavenumber plane. Because $h$ is real,
$\overline{G_{-\lb}}=G_{\lb}$. The bottom elevation spectrum in discrete form
is given by $F^G_{\lb}=\langle G_{\lb} G_{-\lb} \rangle$ and in continuous form
by
\begin{equation}
  F^B(\lb)= \lim _{|\Delta l| \rightarrow 0} \frac{F^G_{\lb}}{\Delta l_x \Delta
  l_y},
\end{equation}
and verifies,
\begin{equation}
  \int_{-\infty}^{\infty} \int_{-\infty}^{\infty} F^B(\lb)\dr l_x {\mathrm
  d}l_y=
  \lim_{L  \rightarrow \infty} \frac{1}{L^2} \int_{-L/2}^{L/2} \int_{-L/2}^{L/2}h^2(x,y) \dr x {\mathrm
  d}y
\end{equation}

 Now that the scene is set, we shall solve for the velocity potential $\phi$ in the frame of reference
moving with the mean current, and use (\ref{NfromPhi}) to estimate the action
spectra density at each successive order. In the course of this calculation,
$\phi$ will appear as the sum of many terms, some of which are secular (these
are the `resonant terms' in Hasselmann's terminology), i.e. with growing
amplitudes in time. Most importantly among these are those that lead to
resonant terms in $N$. All other terms are bounded in time and thus do not
contribute to the long-term evolution of the wave spectrum, i.e. on the scale
of several wave periods, and shall be neglected (see Hasselmann 1962).

\subsection{Zeroth-order solution}
In the moving frame of reference, the governing equations for $\phi_0$ are
identical to those in the fixed frame in the absence of current. The solution
is thus
\begin{equation}
\phi_0=\sum_{\kb,s}\frac{\cosh(k(z+H))}{\cosh(kH)}\Phi^s_{0,\kb}\er^{\ir [\kb
\bcdot \xb -s\sigma t]},
\end{equation}
where the intrinsic frequency $\sigma$ is the positive root of the linear
dispersion relation,
\begin{equation}\label{reldispjfm}
\sigma^2=gk\tanh(kH).
\end{equation}

\subsection{First-order solution}
Surface non-linearity becomes relevant at first order due to a coupling between
the zeroth order solution and current-induced first order terms. Including all
powers of $\eta$, the expansion of the surface boundary condition to order
$\varepsilon^2$ gives, at  $z=0$,
\begin{eqnarray}
\frac{\partial \phi}{\partial t^2}+ g\frac{\partial \phi}{\partial z}& =&-
\zeta \frac{\partial^3 \phi}{\partial t^2 \partial z} - g \zeta
\frac{\partial^2 \phi}{\partial z^2}  -\frac{\partial \zeta}{\partial
t}\frac{\partial^2 \phi}{\partial z \partial t} +\bnabla \phi \bcdot \left(
g\bnabla \zeta - \frac{\partial \bnabla \phi}{\partial t}\right)  -
\frac{\partial \phi}{\partial z} \frac{\partial^2 \phi}{\partial t \partial z}
\nonumber\\& & + c_1'(t) + O(\varepsilon^3)\label{surflin}
\end{eqnarray}

 The equations
at order $\eta$ are
\begin{eqnarray}
  \bnabla^{2}\phi_1+\frac{\partial ^{2}\phi_1}{\partial z^{2}} &
  = & 0 \quad {\mathrm{for}} \quad -H \leq z \leq 0,\label{laplace2} \\
\frac{\partial \phi_1}{\partial z} & = &-h\frac{\partial ^{2}\phi_0}{\partial
z^{2}}+\bnabla \phi_0 \cdot \bnabla h +\frac{\partial h}{\partial t} \qquad
{\mathrm{at}} \qquad z=-H,\label{fond2}
\end{eqnarray}
and, at $z=0$, expansion of (\ref{surflin}) to first order in $\eta$ yields,
\begin{eqnarray}
\frac{\partial ^{2}\phi_1}{\partial t^{2}}&+& g\frac{\partial \phi_1}{\partial
z} = \stackrel{\mathrm{I}} {\overbrace{g\left( \bnabla \phi_0 \bcdot \bnabla
\zeta_1 - \zeta_1 \frac{\partial^2 \phi_0}{\partial z^2} \right)}}
  + \stackrel{\mathrm{II}}
{\overbrace{g\left(\bnabla \phi_1 \bcdot \bnabla \zeta_0 -  \zeta_0
\frac{\partial^2 \phi_1}{\partial z^2}  \right)}} - \stackrel{\mathrm{III}} {
\bnabla \phi_1 \bcdot \frac{\partial \bnabla
\phi_0}{\partial t}}  \nonumber \\
& & \stackrel{\mathrm{IV}} {- \bnabla \phi_0 \bcdot \frac{\partial \bnabla
\phi_1}{\partial t}} \stackrel{\mathrm{V}} {- \left( \frac{\partial
\phi_1}{\partial z} +\frac{\partial \zeta_1}{\partial t} \right)
\frac{\partial^2 \phi_0}{\partial t \partial z}} \stackrel{\mathrm{VI}} {- 2
\frac{\partial \phi_0}{\partial z} \frac{\partial^2 \phi_1}{\partial t
\partial z}}
 \stackrel{\mathrm{VII}} {- \zeta_1 \frac{\partial^3 \phi_0}{\partial t^2 \partial z}}
 - \stackrel{\mathrm{VIII}} { \zeta_0 \frac{\partial^3 \phi_1}{\partial t^2 \partial
 z}}  + NL_1 \nonumber \\ \label{surf2}
\end{eqnarray}
where the terms $NL_1$, not written explicitly (see Hasselmann 1962 eq.
1.11--1.12), are quadratic products of the zeroth-order solution. Since no
gravity waves satisfy both $\sigma=\sigma_1 \pm \sigma_2$ and $\kb=\kb_1 \pm
\kb_2$, $NL_1$ forces a non-resonant wave solution $\phi_1^{\rm nl}$ that will
be neglected because it does not modify our second order wave action balance,
thanks to the choice $\varepsilon < \eta$. The spatially uniform term $c_1'(t)$
has been incorporated into $NL_1$ and is also of second order in the wave
slope, and does not lead to resonances. That term, omitted by Hasselmann
(1962), is responsible for generating microseisms (e.g. Longuet-Higgins
1950\nocite{Longuet-Higgins1950}).

The first-order system of equations is non-linear due to the surface boundary
condition (\ref{surf2}). However, all the right hand side terms of
(\ref{surf2}) are of order $\varepsilon \eta \phi_0$, and thus negligible,
provided that $\phi_1$ is of order $\eta \phi_0$. Without $\partial h/\partial
t$ in (\ref{fond2}) this would be the case, since the other forcing terms are
all proportional to $\eta \phi_0$. However, as suggested by anonymous
reviewers, $\partial h/\partial t$ introduces an external forcing. We thus
first give the solution $(\phi_{1c},\zeta_{1c})$ forced by $\partial h/\partial
t$ only, in the right hand side of (\ref{fond2}). This solution is physically
identical to mean current perturbation caused by the bottom topography and
given by Kirby (1988, his eq. 2.9) for a sinusoidal bottom. With a more general
bottom, it is
\begin{equation}
 \phi_{1c} = \ir  \sum_{\lb} \Ub \bcdot \lb \frac{G_{\lb}}{l \alpha_{\lb}}\left\{\beta_{\lb} \cosh \left[l(z+H)\right]
 + \alpha_{\lb} \sinh \left[l(z+H)\right]\right\} \er^{\ir \lb \bcdot \left(\xb+\Ub
 t\right)},\label{phi1c}
\end{equation}
where
\begin{equation}
 \alpha_{\lb} = \frac{\left(\Ub \bcdot \lb\right)^2}{gl} - \tanh(lh),
\end{equation}
and
\begin{equation}
 \beta_{\lb} = 1-\tanh(lh) \frac{\left(\Ub \bcdot \lb\right)^2}{gl}.
\end{equation}
The corresponding surface elevation oscillations, given by (\ref{Bernoulli}),
are second order in the Froude number $\Frou = U/(gh)^{1/2}$, and 180$^\circ$
out of phase with the bottom oscillations for slow currents when $\alpha<0$
(Kirby 1988, eq. 2.10),
\begin{equation}
 \zeta_{1c} = \sum_{\lb} \frac{\left(\Ub \bcdot \lb\right)^2 G_{\lb}}{g  \alpha_{\lb} \cosh(lh)}
 \er^{\ir \lb \bcdot \left(\xb+\Ub t\right)}.
\end{equation}
From (\ref{phi1c}), the following expression are derived,
\begin{equation}
 \phi_{1c}(z=0) =  \ir  \sum_{\lb} \Ub \bcdot \lb \frac{G_{\lb}}{l \alpha_{\lb} \cosh(lh)}\er^{\ir \lb \bcdot \left(\xb+\Ub t\right)},
\end{equation}
\begin{equation}
 \frac{\partial \phi_{1c}}{\partial z}(z=0)= \frac{\partial \zeta_{1c}}{\partial t}= \ir  \sum_{\lb} \left(\Ub \bcdot \lb\right)^3
 \frac{G_{\lb}}{g \alpha_{\lb} \cosh(lh)}\er^{\ir \lb \bcdot \left(\xb+\Ub
 t\right)}.
\end{equation}
These shall be particularly useful for plugging into (\ref{surf2}).

 We can now obtain the
general solution to our equations (\ref{laplace2})--(\ref{surf2}) by the
following superposition of the previous solution with free and bound (i.e.
non-resonant) wave components, with amplitudes $\Phi^{s}_{1,\kb}$ and
$\Phi^{{\rm si},s}_{1,\kb}$ respectively,
\begin{equation}
\phi_1=\phi_{1c}+\sum_{\kb,s}\left[
\frac{\cosh\left[k(z+H)\right]}{\cosh(kH)}\Phi^{s}_{1,\kb}(t) +
\frac{\sinh\left[k(z+H)\right]}{\cosh(kH)}\Phi^{{\rm si},s}_{1,\kb}(t)
 \right] \er^{\ir \kb \bcdot \xb},
 \label{formphi2}
\end{equation}
where the last two terms corresponds to the solution to the forcing by all the
right hand side terms except for $\partial h/\partial t$. Because $\phi_{1c}$
and $\zeta_{1c}$ are the only terms that may be larger than $\varepsilon \eta
\phi_0$,  all others are neglected in the right-hand side of (\ref{surf2}).

 Substitution of (\ref{formphi2}) in the bottom boundary condition
(\ref{fond2}) yields
\begin{equation}
 \frac{k}{\cosh(kH)} \Phi^{{\rm si},s}_{1,\kb}(t) = -\sum_{\kpb} \frac{\kpb \bcdot \kb}{\cosh(k'H)}\Phi^s_{1,\kpb}G_{\kb-\kpb}\er^{\ir \left[\left(\kb-\kpb\right) \bcdot \Ub -s\sigma'\right]t}.
\end{equation}

Replacing now (\ref{formphi2}) in the surface boundary condition (\ref{surf2}),
yields an equation for $\Phi^{s}_{1,\kb}$. Using $\omega=\sigma+\kb \bcdot \Ub$
and $\omega'=\sigma'+\kb' \bcdot \Ub$, it writes
\begin{equation}\label{oscil2}
\left( \frac{d^2}{dt^2}+\sigma^2 \right)\Phi^s_{1,\kb}(t)= \sum_{\kpb}
M^s(\kb,\kpb)\Phi_{0,\kpb}G_{\kb-\kpb}\er^{\ir \left[\kb \bcdot \Ub
-s\omega'\right]t},
\end{equation}
with
\begin{equation}
M^s(\kb,\kpb)=\left\{ gk-\left[\kb\bcdot \Ub-s\omega'\right]^2 \tanh(kH)
\right\} \frac{\kpb \bcdot \kb}{k}
\frac{\cosh(kH)}{\cosh(k'H)}+M^s_{c1}(\kb,\kpb)
\end{equation}
where $M_{c1}^s$ is given by all the right-hand side terms in (\ref{surf2}) and
thus corresponds to the scattering induced by current and current-induced
surface elevation variations. Anticipating resonance, we only give the form of
$M_{c1}^s=M_{c}^s$ for $\sigma=\sigma'-s \lb \bcdot \Ub$, with $\lb=\kb-\kpb$,
\begin{equation}
M^s_c(\kb,\kpb)=\frac{\{s g^2 \Ub \bcdot \lb(\stackrel{\rm (a)}{\sigma'
\lb\bcdot \kb} + \stackrel{\rm (b)}{\sigma \lb\bcdot \kpb}) -\left(\Ub \bcdot
\lb\right)^2 [\stackrel{\rm (c) }{g^2 \kb \bcdot \kpb} - \stackrel{\rm
(d)}{\overbrace{\sigma \sigma' (\sigma \sigma' +\left(\Ub \bcdot
\lb)^2\right)}} ]\}}{l g^2 \alpha_{\lb} \cosh(lh)},\label{Mc}
\end{equation}
in which the term (a) is given by the term (II) in (\ref{surflin}), (b) is
given by (III) and (IV), (c) is given by (I), and (d) is given by (V)--(VIII).
Because we are first solving the problem to order $\eta$, it is natural that
our solution is a linear superposition of the solutions found by Kirby (1988)
for a single bottom component. Indeed, $M_c(\kb,\kpb)=-4\omega \Omega_c/D$,
with $\Omega_c$ the interaction coefficient of Kirby (1988, eq. 4.22b) and $D$
his bottom amplitude, here $G_l = {\rm i}D/2$.

The solution to the forced harmonic oscillator equation (\ref{oscil2}) is
\begin{equation}  \label{phi2sk}
\Phi^s_{1,\kb}(t)= \sum_{\kpb}M^s(\kb,\kpb)\Phi^s_{0,\kpb}G_{\kb-\kpb}
f_1(\sigma,\lb \bcdot \Ub-s\sigma';t),
\end{equation}
where $\lb = \kb-\kpb$, and the function $f_1$ is defined in Appendix A.

\subsubsection{First order action}
The lowest order perturbation of the wave action by scattering involves the
order $\eta$ covariances
\begin{equation}\label{cov21}
F^\Phi_{1,0,\kb}+F^\Phi_{0,1,\kb}=4 \Real \left(\langle
\Phi_{0,\kb}^{+}\Phi_{1,-\kb}^{-} \rangle\right),
\end{equation}
with $\Real$ denoting the real part. Including only the secular terms, we get
\begin{equation}
F^\Phi_{1,0,\kb}+F^\Phi_{0,1,\kb}= 4\Real \left[\sum_{\kpb} M^+(\kb,\kpb)
\langle \Phi^{+}_{0,\kpb} \Phi^{-}_{0,-\kb} G_{\kb-\kpb}\rangle f_1(\sigma,\lb
\bcdot \Ub -\sigma';t) \er^{\ir \sigma t}\right].
 \label{F21}
\end{equation}
Although this term was assumed to be zero in AH, it is not zero for sinusoidal
bottoms with partially standing waves, and may become significant at resonance
due to the function $f_1$. In uniform conditions, the time evolution of the
wave field requires that the non-stationarity must come into play. Thus $\gamma
\approx \eta$ and the non-stationary term is given by AH (their appendix D),
\begin{equation}
\frac{\partial \left[N_{1,0}^{\mathrm{ns}}(\kb)
+N_{0,1}^{\mathrm{ns}}(\kb)\right]}{\partial t}=-\frac{\partial
N_{0}(\kb)}{\partial t}.\label{nonstat}
\end{equation}
In order to simplify the discussion, we shall briefly assume that there is no
current and that the waves are unidirectional. In that case, $\kb'=-\kb$ and
$M(\kb,\kb')=-g k^2/\cosh^2(kH)$. Replacing (\ref{F21}) in (\ref{nrj_ij}) and
combining it with (\ref{nonstat}) yields the action balance
\begin{equation}
\frac{\partial N_{0,\kb}}{\partial t}= \frac{\partial }{\partial t}
\left[\frac{k}{g \sigma}
\tanh(kH)\left(F^\Phi_{1,0,\kb}+F^\Phi_{0,1,\kb}\right)\right]=\Imag\left(\frac{-4
k^2 \sigma}{2 g \cosh^2(kH)} \langle \Phi^{+}_{0,\kb} \Phi^{-}_{0,\kb} G_{-2k}
\rangle\right),
 \label{E1}
\end{equation}
with $\Imag$ denoting the imaginary part.

For directionally spread random waves, with a current, and a real bottom (e.g.
random or consisting of a finite series of sinusoidal bars), the evaluation of
(\ref{F21}) is not simple. First of all, resonant terms given by $f_1$ only
occur for $\sigma' = \sigma +s \lb \bcdot \Ub$, that is $\omega=\omega'$. Using
$N(\kb)=N_0(\kb) \left[1+O(\eta)\right]$ and taking the limit to continuous
surface and bottom spectra yields
\begin{equation}
\frac{\partial N(\kb)}{\partial t}=  S_1(\kb)= \int_{0}^{2 \upi }  \frac{4 \kb
\bcdot \kpb }{2 g \cosh(kH) \cosh(k'H)} \Imag \left[ Z(\kb,\kpb)\right] \dr
k'_x \dr \theta',
 \label{E3dir}
\end{equation}
with the mixed surface bottom bispectrum $Z$ defined by
\begin{equation}
Z(\kb,\kpb) = \lim_{\Delta \kb \rightarrow\infty}\langle \frac{
\Phi^{+}_{1,\kb} \Phi^{-}_{1,-\kpb} G_{-k-k^\prime}}{\Delta \kb \Delta \theta'}
\rangle,
\end{equation}
with $\kb=k(\cos \theta, \sin \theta)$ and  $\kb^\prime=k(\cos \theta^\prime,
\sin \theta^\prime)$. $Z$ is similar to a classical bispectrum (e.g. Herbers
\etal~2003\nocite{Herbers&al.2003}) with one surface wave amplitude replaced by
a bottom amplitude, and a similar expression is found for a non-zero current.
The action balance (\ref{E3dir}) is generally not closed, and requires a
knowledge of the wave phases that are not available in a phase-averaged model.
The same type of coupling, although due to the large scale topography, also
occurs in the stochastic equations for non-linear wave evolution derived by
Janssen, Herbers \& Battjes (2006\nocite{Janssen&al.2006}).

 The contribution of the mixed bispectrum will
thus be evaluated below, in order to investigate in which cases it may be
neglected or parameterized. It is expected that $S_1$ is generally negligible
because MAHR have neglected $S_1$, and still found a good agreement of the
second order action balance with exact numerical solutions for the wave
amplitude reflection coefficient.

\subsubsection{Second order action}
From the expansion (\ref{Action_exp}), the second order action is $N_2(\kb)=
N_{1,1}(\kb)+N_{0,2}(\kb)+N_{2,0}(\kb)$. The first term can be estimated from
$\phi_1$, using the covariance of the velocity potential amplitudes
(\ref{covarPhi}),
\begin{equation}\label{cov22}
F^\Phi_{1,1,\kb}=2\langle \Phi_{1,\kb}^{+}\Phi_{1,-\kb}^{-}
\rangle.\label{FPHI11}
\end{equation}
Using (\ref{phi2sk}), (\ref{FPHI11}) can be re-written as
\begin{equation}
\frac{F^\Phi_{1,1,\kb}}{\Delta \kb}= 2\sum_{\kpb}\left|M^+(\kb,\kpb)\right|^2
\frac{\langle\left|\Phi^{+}_{0,\kpb}\right|^2\rangle}{\Delta \kpb}
\frac{\langle \left|G_{\kb-\kpb}G_{-\kb+\kpb}\right|^2 \rangle}{\Delta \kb}
 \left|f_1(\sigma,\lb
\bcdot \Ub -\sigma';t)\right|^2
 \Delta \kpb,
 \label{F22}
\end{equation}
Taking the limit of (\ref{F22}) when ${\Delta \kb} \rightarrow 0$,
\begin{eqnarray}\label{PO}
F^{\Phi}_{1,1}(t,\kb) &=&\int_{-\infty}^\infty \int_{-\infty}^\infty
\left|M^+(\kb,\kpb)\right|^2
 F^{\Phi}_{1,1}(\kpb)
F^B(\kb-\kpb) \left|f_1(\sigma,\lb \bcdot \Ub -\sigma';t)\right|^2 \dr k_x' \dr
k_y'.\nonumber \\
\end{eqnarray}
Due to the singularity in $f_1$, and assuming that the rest of the integrand
can be approximated by an anlytical function in the neighbourhood of the
singularity $\omega'=\omega$, which requires both bottom and surface elevation
spectra to be continuous, the integral can be evaluated by using
\begin{equation}
\langle f_1(\sigma,\lb \bcdot \Ub-\sigma';t)f_1(\sigma,-\lb \bcdot
\Ub+\sigma';t)\rangle =\frac{\upi t}{4\sigma^2}\left[\delta
\left(\sigma'-(\sigma+\lb \bcdot \Ub)\right) + O (1)\right]. \label{f1toDirac}
\end{equation}
$\delta$ is the one-dimension Dirac distribution, infinite where the argument
is zero, and such that $\int \delta(x) A(x) \dr x=A(0)$ for any continuous
function $A$. In order to remove that singularity, the argument of $\delta$
maye be re-written as $\omega'-\omega$, making explicit all the dependencies on
$k'$. Evaluation of the $\delta$ function is then performed by changing
integration variables $(k_x',k_y')$ are changed to $(\omega',\theta')$, with a
Jacobian $k' \partial k'/\partial \omega'=k'^2/(k'C_g'+\kpb \bcdot \Ub)$. We
thus have
\begin{eqnarray}\label{PO2}
F^{\Phi}_{1,1}(t,\kb) &=&\frac{\upi t }{2\sigma^2}\int_{0}^{2
\upi}\int_{\omega'} \left|M^+(\kb,\kpb)\right|^2
 F^{\Phi}_{1,1}(\kpb)
\frac{k'F^B(\kb-\kpb) }{C_g'+\kpb \bcdot \Ub} \delta (\omega' - \omega )
d\omega'\dr \theta'+ O(1). \nonumber \\
\end{eqnarray}
When $\omega=\omega'$, the integrand simplifies. $M^s(\kb,\kpb)$ is equal to
$M(\kb,\kpb)$, defined by
\begin{eqnarray}
M(\kb,\kpb)=\frac{g \kb \bcdot \kpb}{\cosh(kH)\cosh(k'H)}+M_c(\kb,\kpb) \equiv
M_b(\kb,\kpb)+M_c(\kb,\kpb),\label{M}
\end{eqnarray}
with $M_c=M_c^+$ given by (\ref{Mc}). Using the (\ref{nrj_ij}) relation between
velocity potential and action, and evaluating the integral over $\omega'$, one
obtains
 \begin{equation}\label{fin2}
N_{1,1}(t,\kb)=\frac{\upi t}{2}\int_{0}^{2 \upi}
M^2(\kb,\kpb)\frac{N_{0,0}(\kpb)}{\sigma \sigma'} F^B(\kb-\kpb)\frac{k'^2}{k'
C_g'+\kpb \bcdot \Ub} d\theta' + O(1).\label{N11}
\end{equation}
Again we note the correspondance with the theory of Kirby (1988, eq. 4.21).
Specifically, one has $M(\kb,\kpb)=-4\omega \Omega_c/D$, with $\Omega_c$ being
Kirby's interaction coefficient.

\subsection{Second order potential and corresponding terms in $N_2$}
In order to estimate the other two terms that contribute to $N_2$, the second
order potential $\phi_2$ must be obtained. It is a solution of
\begin{equation}\label{lap3}
  \bnabla^{2}\phi _{2}+\frac{\partial ^{2}\phi _{2}}{\partial z^{2}}%
=0 \qquad {\mathrm{for}} \qquad -H \leq z \leq 0,
\end{equation}
\begin{equation}
\frac{\partial \phi_2}{\partial z}=-h\frac{\partial ^{2}\phi _{1}}{\partial
z^{2}}-\frac{h^2}{2}\frac{\partial ^{3}\phi _{0}}{\partial z^{3}}+\bnabla \phi
_{1}\cdot \bnabla h +\bnabla (h\frac{\partial \phi _{0}}{\partial z})\cdot
\bnabla h \qquad {\mathrm{at}} \qquad z=-H, \label{bot3JFM}
\end{equation}
that simplifies because odd vertical derivatives of $\phi_0$ are zero at
$z=-H$,
\begin{equation}\label{bottombound3b}
\frac{\partial \phi_2}{\partial z}=-h\frac{\partial ^{2}\phi _{1}}{\partial
z^{2}}+\bnabla \phi _{1}\cdot \bnabla h  \qquad {\mathrm{at}} \qquad z=-H,
\end{equation}
and
\begin{equation}\label{surfbound3}
\frac{\partial ^{2}\phi _{2}}{\partial t^{2}}+g\frac{\partial \phi _{2}}{%
\partial z}=\mathrm{i}\sum_{{\boldsymbol k},s} 2 s \sigma \frac{\partial \Phi_{0,\kb}^{s}}{\partial t}\mathrm{e}^{\mathrm{i}\left( {\boldsymbol
k}\bcdot{\boldsymbol x}-s\omega t\right)} +  {\rm I-VIII} + NL_2 \qquad
{\mathrm{at}} \qquad z=0.
\end{equation}
The terms I--VIII are identical to those in (\ref{surf2}) with $\phi_0$,
$\zeta_0$, $\phi_1$, $\zeta_1$ replaced by  $(\phi_1-\phi_{1c})$,
$(\zeta_1-\zeta_{1c})$, $\phi_2$ and $\zeta_2$, respectively. All other
non-linear terms have been grouped in $NL_2$. In order to yield contributions
to the second order action $N_{2,0}$, terms must correlate with $\phi_0$ to
give second-order terms in $\eta$ with non-zero means. For zeroth order
components with random phases, inspection shows that $NL_2$ do not contribute
to $N_{2,0}$ and will thus be neglected.

The solution $ \phi_2 $ is given by the following form,
\begin{equation}\label{formphi3}
\phi_2=\phi_2^{\mathrm{ns}}+\sum_{\kb,s}\left[
\frac{\cosh(k(z+H))}{\cosh(kH)}\Phi^{s}_{2,\kb}(t) +
\frac{\sinh(k(z+H))}{\cosh(kH)}\Phi^{{\rm si},s}_{2,\kb}(t)
 \right] \er^{\ir \kb
\bcdot \xb}.
\end{equation}
The non-stationarity term $\phi_2^{\mathrm{ns}}$ leads to the action evolution
term (\ref{nonstat}), now assuming $\gamma \approx \eta^2$. Following the
method used at first order, substitution of (\ref{formphi3}) in the bottom
boundary condition (\ref{bottombound3b}) leads to,
\begin{equation}
\Phi^{{\rm si},s}_{2,\kb}(t)=-\sum_{\kb'}\frac{\kb'\cdot
\kb}{k}\frac{\cosh(kH)}{\cosh(k'H)}\Phi_{1,\kb'}^s(t) G_{\kb-\kb'} \er^{\ir
\lb \bcdot \Ub t }.
\end{equation}

After calculations detailed in Appendix B, $\phi_2$ yields the following
contribution to the wave action,
\begin{equation}\label{dE3}
  %\frac{dE_{3,1}(\kb)}{dt}
 N_{2,0}(\kb) + N_{0,2}(\kb)=-\frac{\upi t}{2 } \int_0^{2\upi} M^2(\kb,\kpb)
  F^B(\kb-\kpb)\frac{N_{0}(\kb)}{\sigma \sigma'} \frac{k'^2}{k' C'_g +
  \kpb \bcdot \Ub }\dr  \theta'+ O(1),\label{N20scat}
\end{equation}
in which $\sigma'=\sigma -\lb \bcdot \Ub$, $\sigma'^2=gk'\tanh(kH)$, and
$C_g'=\sigma'(1/2+k'H/\sinh(2k'H))/k'$.

\subsection{Action and momentum balances}
We shall neglect the first order action contribution $N_1$ given by
(\ref{E3dir}). The solvability condition imposed on the action spectrum is that
$N_2$ remains an order $\eta^2$ smaller than $N_0$ for all times. Thus all
secular terms of order $\eta^2$ must cancel. Combining (\ref{nonstat}),
(\ref{N11}), and (\ref{N20scat}) gives
\begin{equation}\label{solvability}
  -\frac{\dr  N_0(\kb)}{\dr t} +\frac{\upi}{2} \int_0^{2\upi} M^2(\kb,\kpb) F^B(\kb-\kpb)\frac{N_{0}(\kpb)-N_{0}(\kb)}{\sigma \sigma'} \frac{k'^2}{k' C'_g +
  \kpb \bcdot \Ub }\dr  \theta'
  \end{equation}
Since $N_2$ and $N_1$ remain small,  $N(\kb)=N_0(\kb)
\left[1+O(\eta^2)\right]$, and one has,
\begin{equation}\label{nrjbalance1}
\frac{\dr  N(\kb)}{\dr t}=S_{\mathrm{bscat}}(\kb)\label{action_balance},
\end{equation}
with the spectral action source term,
\begin{equation}\label{nrjbalance2}
 S_{\mathrm{bscat}}(\kb) =
\frac{\upi}{2}\int_{0}^{2 \upi} \frac{k'^2 M^2(\kb,\kpb)}{\sigma
\sigma'\left(k' C'_g + \kpb \bcdot \Ub \right)}F^B(\kb-\kpb) \left[N(\kpb)-
N(\kb)\right] \dr \theta',
\end{equation}
where  $\sigma'=\sigma+\lb \bcdot \Ub$ and $\kb=\kpb+\lb$. This interaction
rule was already given by Kirby (1988). The only waves that can interact share
the same absolute frequency $\omega = \sigma + \kb \bcdot \Ub= \sigma'+ \kpb
\bcdot \Ub$. For a given $\kb$ and without current, the resonant $\kpb$ and
$\lb$ lie on circles in the wavenumber plane (see AH). The current slightly
modifies this geometric property. For $U<<C_g$ the circles become ellipses
(Appendix C).

For a given value of $\omega$, one may obtain the source term integrated over
all directions,
\begin{eqnarray}\label{intS}
S_{\mathrm{bscat}}(\omega)& = &\int_{0}^{2 \upi} k S_{\mathrm{bscat}}(\kb)
\frac{\partial k}{\partial \omega} \dr \theta \\
&=&\int_{0}^{2 \upi}\int_{0}^{2 \upi}\frac{\upi}{2}\frac{k^2 k'^2
M^2(\kb,\kpb)F^B(\kb-\kpb) }{\sigma \sigma' \left(k' C'_g + \kpb \bcdot \Ub
\right)\left(k C_g +
  \kb \bcdot \Ub \right)}\left[ N(\kpb)-N(\kb) \right] d\theta'\dr
\theta \nonumber \\
&=&\int_{0}^{2 \upi} \int_{0}^{2 \upi}
\frac{\upi}{2}\frac{M^2(\kb,\kpb)F^B(\kb-\kpb)}{\sigma \sigma'} \left[
\frac{k^2 N(\omega,\theta')}{k C_g + \kb \bcdot \Ub}-\frac{k'^2
N(\omega,\theta)}{k' C'_g + \kpb \bcdot \Ub} \right] d\theta'\dr \theta.
\nonumber
\end{eqnarray}
This expression is anti-symmetric, multiplied by -1 when $\theta$ and $\theta'$
are exchanged. Thus $S_{\mathrm{bscat}}(\omega)$ is a substraction of two equal
terms, so that for any bottom and wave spectra $S_{\mathrm{bscat}}(\omega)=0$.
In other words, the `source term' is rather an `exchange term', and conserves
the wave action at each absolute frequency. This conservation is consistent
with the general wave action conservation theorem proved by Andrews \& McIntyre
(1978\nocite{Andrews&McIntyre1978b}), which states that there is no flux of
action through an unperturbed boundary (here the bottom). It also appears that
$\omega$ and $\theta$ are natural spectral coordinates in which the scattering
source term takes a symmetric form. Finally, we may consider the equilibrium
spectra that satisfy $S_{\mathrm{bscat}}(\kb)=0$ for all $\kb$. Without
current, an equilibrium exists when either $N(\omega,\theta)$ or $N(\kb)$ is
isotropic. With current, the scattering term is uniformly zero if and only if
the spectral densities in $\kb$-space, $N(\kb)$, are uniform along the curves
of constant $\omega$.

The source term $ S_{\mathrm{bscat}}$ may also be re-written in a form
corresponding to that in AH, which now appears much less elegant,
\begin{equation}\label{nrjbalance2bis}
 S_{\mathrm{bscat}}(\kb) =
\int_{0}^{2 \upi} K(k,k',H) F^B(\kb-\kpb) \left[ N(\kpb)-N(\kb) \right]
d\theta',
\end{equation}
with
\begin{equation}
 K(k,k',H)=\frac{\upi k'^2 M^2(k,k')}{2 \sigma \sigma' \left(k' C'_g +
  \kpb \bcdot \Ub \right)}=\frac{4\upi\sigma k
k'^3 \cos^2(\theta-\theta')
\left[1+O(\Frou)\right]}{\sinh(2kH)\left[2k'H+\sinh(2k'H)\left(1+2 \kpb \bcdot
\Ub /\sigma' \right)\right]}.
\end{equation}

One may wonder how large is the current-induced scattering represented by
$M_c$, our eq. (\ref{Mc}),  compared to the bottom-induced scattering
represented by $M_b$. Since $\sigma'=\sigma + (\Ub \bcdot \lb)$, the (a) and
(b) terms in the numerator $M_c$ almost cancel for small Froude numbers, and
the (a)+(b) part is of order $Fr^2$. Thus $M_c$ is generally an order $Fr^2$
smaller than $M_b$. For $\kb$ and $\kpb$ in opposite directions (i.e.
back-scattering), the (a)+(b) part is even smaller, of order $g^2 (\Ub \bcdot
\lb)^3 \lb \bcdot \kb$, and exactly zero in the long wave limit $lH \ll 1$.
Thus, for back-scattering, the numerator in $M_c$ is itself of the order of
(c), i.e. $g^2 (\Ub \bcdot \lb)^2 \kb \bcdot \kpb$. Interestingly (c) formally
comes from the modulations of the surface elevation $\zeta_{1c}$ so that the
$O(\Frou^2)$ elevation modulation is at least as important as the $O(\Frou)$
current modulation for this back-scattering situation. In that case, $M_c$ is
of the order of $M_b \cosh(kH) \cosh(k'H) (\Ub \bcdot \lb)^2 /[g l \alpha_l
\cosh(lH)]$. The relative magnitudes of $M_b$ and $M_c$ thus depend on
$\Frou(l)= (\Ub \bcdot \lb)/[g l \tanh(l H)]^{1/2}$ that appears in $(\Ub
\bcdot \lb)^2/(g l \alpha_l)=\Frou^2(l)/[\Frou^2(l)-1]$. This $l$-scale Froude
number may be formally close to 1, and thus $M_c$ may be larger than $M_b$.
However, scattering is limited by blocking as no scattered waves can propagate
when $Cg'<\Ub \bcdot \kpb / k'$. In the long wave limit, $\Frou(l)=\Frou$ and
for $(1-\Frou) \ll 1$, one has $M_c > M_b$. For oblique scattering, the (a)+(b)
term may dominate the numerator of $M_c$ and the situation is more complex.
Nevertheless, for Froude numbers typical of continental shelf situations, say
$0<\Frou<0.4$, $M_c$ may be neglected in most situations since its $O(Fr^2)$
correction corresponds to only a few percent of the reflection. Obvious
exceptions are cases in which $M_b$ is zero, such as when $\kb$ and $\kpb$ are
perpendicular.

 Finally, we may also write the evolution equation for the wave
pseudo-momentum ${\mathbf M}^w = \rho_w g \int \kb N(\kb)\dr \kb$  (see Andrews
\& McIntyre 1978), where $\rho_w$ is the density of sea water. Introducing now
the slow medium and wave field variations given by Kirby (1988), that do not
interfere with the scattering process, except by probably reducing the
surface-bottom bispectrum $Z$, one obtains an extension of the equation of
Phillips (1977\nocite{Phillips1977})
\begin{equation}
   \frac{\partial M_\alpha^w}{\partial t} + \frac{\partial}{\partial
x_\beta}\left[ (U_{\beta}+ C_{g \beta})  M_\alpha^w \right]  =
-\tau^{\mathrm{bscat}}_\alpha -M_\beta^w \frac{\partial U_\beta}{\partial
x_\alpha} - \frac{M_\alpha^w}{k_\alpha} \frac{k \sigma}{\sinh 2kD}
\frac{\partial D}{\partial x_\alpha}, \label{wavemom}
\end{equation}
with the dummy indices $\alpha$ and $\beta$ denoting dummy horizontal
components, and the scattering stress vector,
\begin{equation}
 {\tau}^{\mathrm{bscat}}=-\rho_w g \int \kb S_{\mathrm{bscat}} \dr \kb.\label{Tbscat}
\end{equation}
This stress has dimensions of force per unit area, and corresponds to a force
equal to the the divergence of the wave pseudo-momentum flux. Based on the
results of Longuet-Higgins (1967\nocite{Longuet-Higgins1967}) and Hara \& Mei
(1987), this force does not contribute to the mean flow equilibrium with a
balance of the radiation stresses divergence by long waves (or wave set-up in
stationary conditions), contrary to the initial proposition of Mei (1985). This
force is thus a net flux of momentum through the bottom, arising from a
correlation between the non-hydrostatic bottom pressure and the bottom slope.
That force is likely related to the pressure under partial standing waves
locked in phase with the bottom undulations. Although the part $M_c$ of the
coupling coefficient $M$ given by (\ref{M}) is formally due to scattering by
the current modulations $\bnabla \phi_{1c}$, and associated surface
fluctuations $\zeta_{1c}$, it should be noted that these motions and related
pressures are correlated with the bottom slope in the same way as the part
represented by $M_b$. Thus both terms contribute to this force
${\tau}^{\mathrm{bscat}}$ which acts on the bottom and not on the mean flow.

\section{Wave scattering in two dimensions}
Before considering the full complexity of the 3D wave-bottom scattering in the
presence of a current, we first examine the behaviour of the source term in the
case of 2D sinusoidal seabeds. Although the bottom spectrum is not continuous
along the $y$-axis, continuity in $x$ is sufficient for the use of
(\ref{f1toDirac}) and the source term can be applied, after proper
transformation to remove these singularities. MAHR have investigated the
applicability limits of the source term with $U=0$. They proved that for small
bottom amplitudes the source term yields accurate reflection estimates, even
for localized scatterers, and verified this with test cases. It is thus
expected that this also holds for $U \neq 0$.

 \subsection{Wave evolution equation in $2$D}
 We consider here a steady wave field in two dimension with incident and
 reflected waves propagating along the $x$-axis. We shall consider in particular the
 case of $m$ sinusoidal bars of amplitude $b$ and height $2b$, with a wavelength $2\upi /l_0$.
 The bottom elevation is thus
 \begin{eqnarray}\label{hsin}
h(x) & = & b \sin(m l_0 x) \quad {\mathrm{for}} \quad 0 < x < L \label{sinbot}\\
h(x) & = & 0 \quad {\mathrm{otherwise}}. \nonumber
\end{eqnarray}
Such a bottom is shown in figure 1 for $m=4$. This form is identical to that of
the bottom profile chosen by Kirby (1988) but differs, for $0<x<L$, by a
$\upi/2$ phase shift from the bottom profile chosen by Mei (1985). The bottom
spectrum is of the form
 \begin{equation}
F^B(l_x,l_y) = F^{B2D}(l_x) \delta(l_y),\label{FB2D}
\end{equation}
and for the particular bottom given by (\ref{sinbot}),
 \begin{equation}
F^{B2D}(l_x) = \left(\frac{1}{2 \upi} \int_{-\infty}^{\infty} h(x)\er^{-\ir  l
x} \dr x\right)^2= \frac{2 b^2 l_0^2}{\upi L}\frac{\sin^2(l
L/2)}{(l_0^2-l^2)^2},\label{FB2Dsin}
\end{equation}
with
 \begin{equation}
F^{B2D}(\pm l_0) = \frac{m b^2}{4 l_0} = \frac{b^2 L}{8 \upi}.\label{specatres}
\end{equation}
Note that this is a double-sided spectrum, with only half of the bottom
variance contained in the range $l_x>0$. For a generic bottom, for which $h(x)$
does not go to zero at infinity, the spectrum is obtained using standard
spectral analysis methods, for example, from the Fourier transform of the
bottom auto-covariance function (see MAHR). In that case $F^{B2D}$ is
equivalent to a Wigner distribution (see e.g. Ryzhik \etal~1996).

First, replacing (\ref{FB2D}) in (\ref{nrjbalance1}) removes the angular
integral in the source term. Taking $\kb=(k_x,k_y)$, we have $l_y = k_y-k'_y=k
\sin \theta-k'\sin \theta' $, thus ${\rm d}l_y$= $-k'_y cos \theta' {\rm
d}{\theta'}$, and
\begin{equation}
S_{\mathrm{bscat}}\left(\kb,x\right) =\frac{\upi k' M^2(k,k')
F^{B2D}(\kb-\kpb)}{2 \sigma \sigma' \left|\cos \theta'\right| \left(k' C'_g +
\kpb \bcdot \Ub \right)}\left[N(\kpb)-N(\kb) \right].
\end{equation}
Second, assuming now that waves propagate only along the $x$-axis, the wave
spectral densities are of the form
 \begin{equation}
N(k_x,k_y) = N(k_x,k_y) \delta(k_y)=N^{2D}(k)\delta(\theta-\theta_0)/k,
\end{equation}
with $\theta_0=0$ for $k_x >0$ and $\theta_0=\upi$ for $k_x < 0$.
 Integrating over $\theta$ removes the singularities on $k_y$, and assuming a
steady state one obtains
 \begin{equation}
 \label{energybalance1D}
 \left[\frac{k_x}{k} C_g + U_x \right]\frac{\partial N^{2D}}{\partial
 x}\left(k_x,x\right)=S_{\mathrm{bscat}}^{2D}\left(k_x,x\right),\label{Nevol_kx}
 \end{equation}
with
\begin{equation}
S_{\mathrm{bscat}}^{2D}\left(k_x,x\right)=\frac{\upi k' M^2(k,k')
F^{B2D}(k_x-k'_x)}{2 \sigma \sigma'\left(k_x' C'_g + k'_x U_x \right)}
\left[N^{2D}(k_x',x)-N^{2D}(k_x,x)\right].
\end{equation}

Although the present theory is formulated for random waves, there is no
possible coupling between waves of different frequencies. Mathematically, it is
possible to take the limit to an infinitely narrow wave spectrum, such that,
$N^{2D}(k,x)=N(x) \delta(\omega-\omega_0)+N'(x) \delta(\omega'-\omega'_0)$ with
$k_{0x} >0$ and $k'_{0x} <0$. Using $\partial \omega/\partial k = C_g +  k_x
U_x/\left|k_x\right|$, the resulting evolution equation is, omitting the 0
subscripts on $k$ and $k'$,
\begin{eqnarray}
& &\left[\frac{k_x}{k} C_g  + U_x \right] \frac{\partial N}{\partial
 x} \nonumber\\
 & &=\frac{\upi  M^2(k,k') F^{B2D}(k_x-k'_x)}{2
\sigma \sigma' }\left[\frac{
 k N'}{k C_g + k_x U_x}
 -  \frac{k' N}{k' C'_g + k'_x U_x}\right],\nonumber\label{Nevol_mono}\\
\end{eqnarray}
with a similar equation for $N'$ obtained by exchanging $C_g$ and $C_g'$, and
$k'$ and $k$, from which it is easy to verify that the total action is
conserved.

 The stationary evolution equation (\ref{Nevol_kx}) only couples two wave
components $N(k)$ and $N(k')$. For a uniform mean depth $H$, and uniform bottom
spectrum $F^B$, as considered here, we thus have a linear system of two
differential equations, that may be written in matrix form for any $k>0$,
\begin{equation}
\frac{\dr}{\dr x} \left(\begin{array}{c}
  N(k) \\
  N(k')\end{array}
\right)  = q \sfbsQ  \left(\begin{array}{c}  N(k) \\
  N(k')\end{array}\right),
\end{equation}
with
\begin{equation}
q=\frac{\upi M^2(k,k')F^{B2D}(l)}{2 \sigma \sigma' Cg Cg'}\label{defq}
\end{equation}

 Defining
$l=k_x-k'_x$, the action advection velocities $V'=C_g'+k'_x U_x$ and $V=C_g+k_x
U_x$, the terms of the non-dimensional matrix $\sfbsQ$ are given by
\begin{eqnarray}
(\sfbsQ)_{1,1} = -\frac{C_g C_g'}{ V^2} &\quad {\rm and} \quad&
 (\sfbsQ)_{1,2}  = \frac{C_g C_g'}{ V V'}, \nonumber \\
(\sfbsQ)_{2,1}  = -\frac{C_g C_g'}{V'^2} &\quad {\rm and} \quad& (\sfbsQ)_{2,2}
=\frac{C_g C_g'}{V V'},
\end{eqnarray}
where $(\sfbsQ)_{i,j}$ is the $i^{\rm th}$ row and $j^{\rm th}$ column term of
$\sfbsQ$. The general solution is thus
\begin{equation}
\left(\begin{array}{c}
  N(k,x) \\
  N(k',x)\end{array}
\right)  =   \er^{q \sfbsQ x}  \left(\begin{array}{c}  N(k,0) \\
  N(k',0)\end{array}\right).
\end{equation}
The matrix exponential is classically the infinite series $\sum_{n=0}^\infty
\left( q \sfbsQ\right)^n/n!$, in which matrix multiplications are used. The
reflection coefficient for the wave action is found using the boundary
condition expressing the absence of incoming waves from beyond the bars,
$N(k',L)=0$, giving,
\begin{equation}
R_N  =\frac{N(k',0)}{N(k,0)}=  -\left(\er^{q \sfbsQ
L}\right)_{2,1}/\left(\er^{q \sfbsQ L}\right)_{2,2}.
\end{equation}
A reflection coefficient for the modulus of the wave amplitude predicted by the
source term is thus,
\begin{equation}
R_S = \left[\frac{\sigma' N(-k',0)}{\sigma
N(k,0)}\right]^{1/2}=-\left\{\sigma'\left(\er^{q \sfbsQ
L}\right)_{2,1}/\left[\sigma \left(\er^{q \sfbsQ
L}\right)_{2,2}\right]\right\}^{1/2} \label{Krsbscat}.
\end{equation}
The spatial variation of the amplitudes may be linear, oscillatory, or
exponential, depending on whether the determinant of $\sfbsQ$, is zero,
negative or positive, respectively. That determinant is $C_g^2  C_g'^2
(V'-V)(V'^2+3V V' + 4 V^4)/V^4V'^3$, which is always of the sign of $V'-V$.

\subsection{Analytical solution for $U=0$}
\begin{figure}
\centerline{\includegraphics[width=\textwidth]{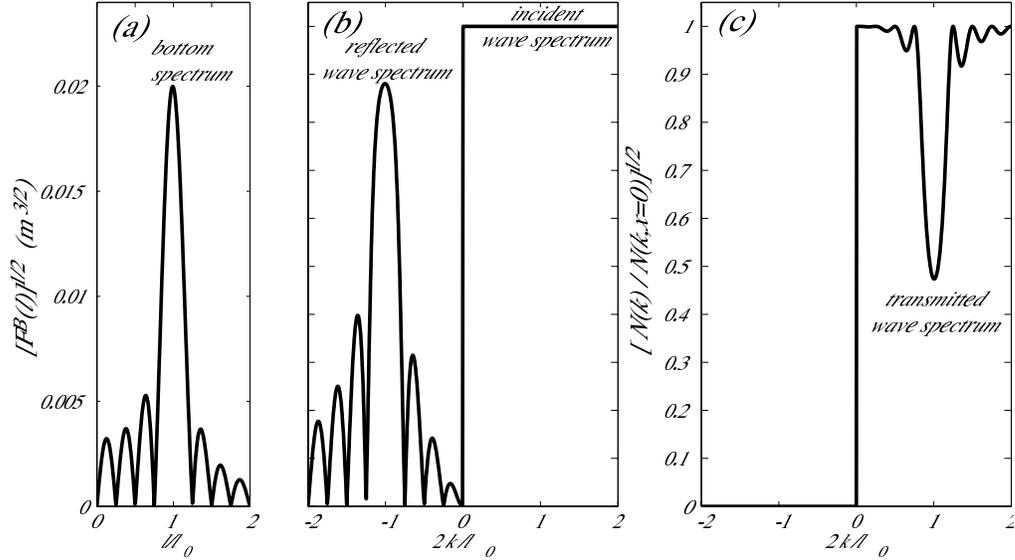}}
 \caption{Bottom spectrum and evolution of a surface wave spectrum along a field of sinusoidal bars for
 $U=0$, $b=0.05$~m, $H=0.156$~m, so that $\eta=b/H=0.32$, and $l_0=2 \upi$, $n=4$, so that $L=4~m$ (bottom
 shown in figure 1).
 (a) square root of the bottom spectrum, (b) and (c) normalized square root wave spectrum upwave (at $x<0$) and downwave (at $x>L$) of the bars, respectively.
  The incident spectrum ($k>0$ at $x=0$) is specified to be  white (unform in wavenumbers).}
 \label{Sp_evol}
\end{figure}

  In the absence of a mean current, $k'=-k$, and
\begin{eqnarray}
(\sfbsQ)_{1,1}=(\sfbsQ)_{1,2}=-(\sfbsQ)_{2,1}=(\sfbsQ)_{1,1}=1,
\end{eqnarray}
Thus $\sfbsQ^2=0$ so that its exponential is only the sum of two terms, $\er^{q
\sfbsQ x}= \left(\sfbsI+q \sfbsQ\right) x$, where $\sfbsI$ is the identity
matrix. The solution to (\ref{Nevol_mono}) is simply,
 \begin{eqnarray}
 N(k,x)& = & N(k,0) \left[\frac{- q \left(x-L\right) +1}{1+q L}\right] \\
 N(-k,x) &= & N(k,0)\left[\frac{-  q \left(x-L\right) }{1+q
 L}\right].
 \end{eqnarray}
An example of spatial variation of the wave spectrum from $x=0$ to $x=L$ is
shown in Figure \ref{Sp_evol}, for $U=0$, and a uniform (white) incident
spectrum. The reflected wave energy (at $k<0$ in figure
\ref{Sp_evol}.\textit{a}) compensates the loss of energy in the transmitted
spectrum (at $k>0$ in figure \ref{Sp_evol}.\textit{b}).
%%%%%%%%%%%%%%%%%%%%%%%%%%%%%%%%%%%%%%%%%%%%%%%%%%%%%%%%%%%%%%%%%%%%%%%%%%%%%%%%%%%%%
%%%%%%%%%%%%%%%%%%%%%%%%%%%%%%%%%%%%%%%%%%%%%%%%%%%%%%%%%%%%%%%%%%%%%%%%%%%%%%%%%%%%%%

For $k=l/2$, in the limit of small bar amplitudes, and replacing
(\ref{specatres}) in (\ref{Krsbscat}) yields
\begin{equation}
R_{S} = \left(q L\right)^{1/2} + O (qL) = \frac{k^2 b L}{2 kH + \sinh (2kH)}+ O
(qL)
\end{equation}
which is identical to Mei's (1985) equation (3.21)--(3.22) for exact resonance,
in the limit of $qL \ll 1$, and also converges to the result of Davies \&
Heathershaw (1984) for that same limit. For large bar amplitudes, the
reflection is significant if the bars occupy a length $L$ longer than the
localization length $1/q$. However, the reflection coefficient for the wave
amplitude only increases with $L$ as $\left[q L / (1+q L)\right]^{1/2}$, which
is slower than the exponential asymptote given by Mei (1985) for sinusoidal
bars, and predicted by (Belzons \etal~1988) from the lowest-order theory
applied to a random bottom. The present inclusion of the correlations of
second-order and zeroth order terms may be thought as the representation of
multiple reflections that tend to increase the penetration length in the random
medium.

\begin{figure}
\centerline{\includegraphics[width=0.7\textwidth]{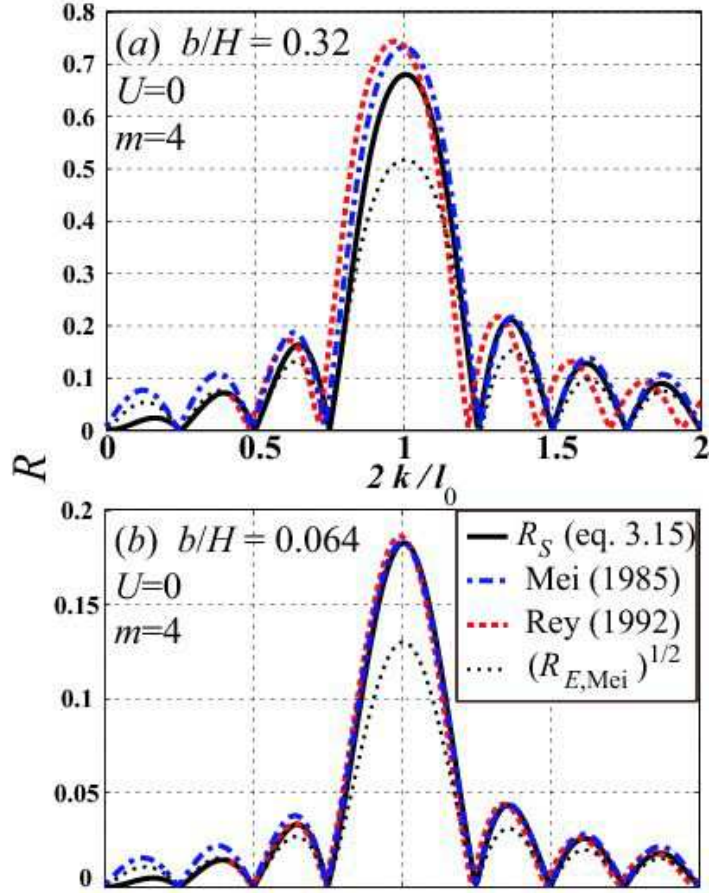}}
 \caption{Reflection coefficients for the wave amplitudes for
 $U=0$, $H=0.156$~m, $l_0=2 \upi$, $n=4$. In (a) $b=0.05$ so that $\eta=b/H=0.32$,
 corresponding to one of the experiments of Davies \& Heathershaw (1984),
 and in (b), $b=0.01$, so that $\eta=b/H=0.064$.}
 \label{CompMei}
\end{figure}

A deeper understanding of this question is provided by the comparison of
numerical estimations of the reflection coefficients for the wave amplitudes
$R$. A benchmark estimation for linear waves is provided by the step-wise model
of Rey (1995) using integral matching conditions for the free propagating waves
and three evanescent modes at the step boundaries. This model is known to
converge to the reflection coefficents given by an exact solution of Laplace's
equation and the boundary conditions, in the limit of an infinite number of
steps and evanescent modes. Calculations are performed here with 70 steps and 3
evanescent modes. These numbers are chosen because a larger number of steps or
evanescent modes gives indistinguishable results in figure \ref{CompMei}.
Results of the benchmark model are in good agreement with the measurements of
Davies \& Heathershaw (1984), except for wave components for which the
reflection over the beach, not included in the model, is comparable to the
reflection over the bars. An analytical expression $R_{\mathrm{Mei}}$ is given
by Mei (1985). $R$ for the present second order theory is given by $R_{S}$
(\ref{Krsbscat}).

We further compare these estimates to the reflection coefficient
$R_{E,{\mathrm{Mei}}}$ that is deduced from the energy evolution given by Hara
\& Mei (1987\nocite{Hara&Mei1987}), using the approximate solutions of Mei
(1985, his equations 3.8--3.23). One may prefer to reformulate the energy
evolution from the amplitude evolution equations of Kirby (1988) because he
used a continuous water depth $h=\sin (ml_0)$, instead of Mei's $h=\cos(ml_0)$
which is discontinuous at $x=0$ and $x=L$\footnote{Such a discontinuous bottom
has a markedly different spectrum at low and high frequencies. The present
theory, confirmed by calculations with Rey's (1995) numerical model, yield very
different reflection coefficients for waves much shorter and much longer than
the resonant waves}. Yet both Mei's and Kirby's equations lead to the same
energy exchange between the incident and reflected components. Using Mei's
(1985) notations, the amplitudes of the incident waves, reflected waves, and
bottom undulations are $A=2 \sigma \Phi^{+}_{0,\kb}/g$, $B=2\sigma
\Phi^{-}_{0,\kb}/g$, and $D=-2\ir  G_{-2k}$, and the `cut-off' frequency is
\begin{equation}
\Omega_0=\frac{\sigma k D}{2 \sinh(2kH)}.\label{Om0}
\end{equation}
The energy evolution of waves propagating over sinusoidal bars along the
$x$-axis is given by Hara \& Mei (1987\nocite{Hara&Mei1987}). The reflected
wave energy ${BB^\star}/{2}$ should be a solution of
\begin{equation}
\frac{\partial }{\partial t} \left(\frac{BB^\star}{2}\right) -  C_g
\frac{\partial }{\partial x} \left(\frac{BB^\star}{2}\right) = \Real \left(\ir
\Omega_0 B^\star A \right),
 \label{EMei}
\end{equation}
where $B^\star$ denotes the complex conjugate of  $B$. This is identical to
(\ref{E1}) for a \textit{monochromatic}  bottom except that the imaginary part
replaced by a real part.

Equation (\ref{EMei}) yields a corresponding energy reflection coefficient,
given by the fraction of energy lost by the incoming waves,
\begin{equation}
R_{E,{\rm Mei}} =-\frac{1}{C_g}\int_0^L \Real \left(\ir  \Omega_0 B^\star A
\right) \dr x.
\end{equation}
Simple analytical expressions can be obtained at resonance, where Mei's (1985)
eq. (3.20)--(3.21) give,
\begin{equation}
\frac{AB^\star}{A^2(0)}=\frac{-\mathrm{i} \sinh
\left(2\tau(1-x/L)\right)}{2\cosh^2\tau}
\end{equation}
with $\tau = \Omega_0 L / C_g$, so that
\begin{equation}
R_{E,{\rm Mei}} = \frac{\cosh 2\tau -1}{4 \cosh^2\tau} = \frac{1}{2} \tanh^2
\tau = \frac{1}{2} R_{{\mathrm{Mei}}}^2,\label{RES1Mei}
\end{equation}
and
\begin{equation}
R_{E,{\rm Mei}} ={2}^{-1/2} R_{{\mathrm{Mei}}}.\label{RS1Mei}
\end{equation}
 It is not surprising that the energy transfer thus computed differs from
the energy computed from the amplitude evolution equations. This is typical of
small perturbation methods, and was discussed by Hasselmann
(1962\nocite{Hasselmann1962}), among others. Yet, it is remarkable that the
ratio of the two is exactly one half. The transfer of energy given by $\ir
\Omega_0 B^\star A$ in (\ref{EMei}) thus correspond to an amplitude reflection
coefficient $R_{E,{\rm Mei}}$ that is smaller by a factor $2^{-1/2}$, at
resonance, compared to $R_{\mathrm{Mei}}$ (figure 3). This underprediction of
the the reflexion of the energy by (\ref{RES1Mei}) also has consequences for
the analysis and calculation of wave set-up due to wave group propagation over
a reflecting bottom. Indeed, the estimation of the scattering stress
(\ref{Tbscat}), that contribute to the driving of long waves, was analyzed by
Hara \& Mei (1987) using a calculation similar to (\ref{RES1Mei}), which is a
factor 2 too small. This may explain, in part, their under-prediction of the
observed elevation of the long wave travelling with the incident wave group.
However, the present theory, compared to that of Hara \& Mei (1987), is limited
to small bar amplitudes, and fails to reproduce their observation of the
transition from oscillatory to exponential decay in the spatial evolution of
the wave amplitude.

\subsection{Effects of wave and bottom relative phases}
The energy exchange coefficient given by the source term always gives energy to
the least energetic components (in the absence of currents), and thus the
energy evolution is monotonic. The action source term (\ref{E1}) of order
$\eta$, that was neglected so far, may have any sign, and thus lead to
oscillatory evolutions for the wave amplitudes, as predicted by Mei (1985) and
observed by Hara \& Mei (1987). At resonance, and for $U=0$, it can be seen
that the first-order energy product $\Phi^{+}_{0,\kb} \Phi^{-}_{0,\kb} G_{-2k}$
in (\ref{E1}) is equal to $\ir  A B^\star D /8$, in the limit of a large number
of bars. Based on Mei's (1985) approximate solution, in the absence of waves
coming from across the bars, this quantity is purely real so that its imaginary
part is zero and the corresponding reflection coefficient $R_{S1}$ is zero. For
$U\neq 0$ this property remains as can be seen by replacing Mei's (1985)
solution with Kirby's (1988). However, similar correlation terms were also
neglected in the second order energy (Appendix B), so that the oscillations of
the amplitude across the bar field, observed by Hara and Mei (1987) may occur
due to terms of the same order as the scattering source term, including
interactions of the sub-harmonic kind (Guazzelli \etal~1992). Further, the
bottom-surface bispectrum in $S_1$ may become significant if there is a large
amount of wave energy coming from beyond the bars. This kind of situation, e.g.
due to reflection over a beach, was discussed by Yu \& Mei
(2000\nocite{Yu&Mei2000}).

%%%%%%%%%%%%%%%%%%%%%%%%%%%%%%%%%%%%%%%%%%%%%%%%%%%%%%%%%%%%%%%%%%%%%%%%%%%%%%%%%%%%%
%%%%%%%%%%%%%%%%%%%%%%%%%%%%%%%%%%%%%%%%%%%%%%%%%%%%%%%%%%%%%%%%%%%%%%%%%%%%%%%%%%%%%%
In the absence of such a reflection, and away from resonance but for small
values of the scattering strength parameter $\tau=(qL)^{1/2}=\Omega_0 L/C_g$,
the imaginary part of $\Phi^{+}_{0,\kb} \Phi^{-}_{0,\kb} G_{-2k}$ is an order
$(qL)^{1/2}$ smaller than the real part and thus contributes a negligible
amount to the reflection.

\subsection{Source term and deterministic results for sinusoidal bars}
For large bar amplitudes, such as $\eta=b/H=0.32$ (figure 3.\textit{a}), all
theories with linearized bottom boundary conditions fail to capture the shift
of the reflection pattern to lower wavenumbers. This effect was discussed by
Rey (1992\nocite{Rey1992}), and attributed to the non-linear nature of the
dispersion relation and the rapid changes in the water depth. Reflection
coefficients are still relatively well estimated. For these large amplitudes
Mei's (1985) approximate solution is found to be more accurate at resonance
compared to the source term. As expected from MAHR and proved here,
$R_{{\mathrm{Mei}}}$ and $R_{S}$ become identical as $\eta=b/H$ goes to zero
(figure \ref{CompMei}.\textit{b}). This fact provides a verification that the
first order scattering term $S_1$ is different from Hara and Mei's (1987)
energy transfer term, and only accounts for a small fraction of the reflection,
a fraction that goes to zero as $\eta \rightarrow 0$. It is also found that for
all bottom amplitudes,  the source term expression provides a simple and
accurate solution away from resonance.
% , for which Mei's (1985) approximate solutions (see the
%sidelobes in figure 3).

\begin{figure}
\centerline{\includegraphics[width=0.7\textwidth]{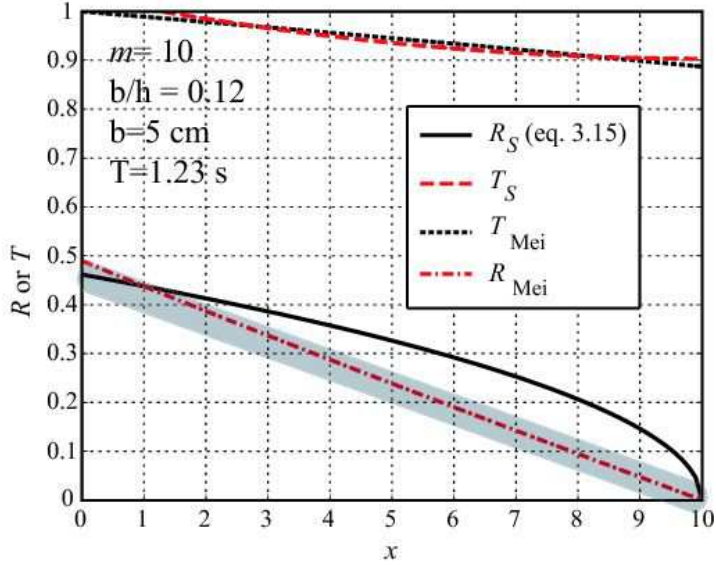}}
 \caption{Spatial evolution of the incident and reflected wave amplitudes
 represented by transmission ($T$) and reflection ($R$) coefficients, in the
 near-resonant case
 $U=0$, $H=0.156$~m, $l_0=2 \upi$~m$^{-1}$, $m=10$ bars, $b=5$~cm, $\eta=b/H=0.12$ and with a wave period $T=1.23$~s. This
 situation corresponds to one of the experiments of Davies \& Heathershaw (1984),
 and their measurements lie in the shaded area.}
 \label{CompDH10}
\end{figure}

Nevertheless, the scattering source term cannot give an accurate description of
the spatial variation of the wave amplitude over a deterministic bottom, as
shown in figure \ref{CompDH10}. This is related to the fact that, in MAHR, the
present reflection coefficient was obtained from the theory of Pihl
\etal~(2002) after averaging over the auto-correlation scale of the bottom
topography. The present theory can only provide an accurate description of the
spatial evolution of the wave field over scales larger than this bottom
auto-correlation distance.
%%%%%%%%%%%%%%%%%%%%%%%%%%%%%%%%%%%%%%%%%%%%%%%%%%%%%%%%%%%%%%%%%%%%%%%%%%%%%%%%%%%%%
%%%%%%%%%%%%%%%%%%%%%%%%%%%%%%%%%%%%%%%%%%%%%%%%%%%%%%%%%%%%%%%%%%%%%%%%%%%%%%%%%%%%%%

 \subsection{Effects of currents}
A prominent feature of solutions with current is the modification of the
resonant condition from $k=k'$ and $l=2k$, to $\sigma'=\sigma + l U$ and
$l=k+k'$, discussed in detail by Kirby (1988). This shift was verified in the
laboratory by Magne, Rey \& Ardhuin (2005\nocite{Magne&al.2005b}). The
magnitude of the resonant peak is also largely enhanced for waves against the
current, due to a general conservation of the action fluxes and the variation
in the action transport velocity, from $C_g+U$ for the incident waves, to
$C_g'-U$ for the reflected waves. Further, the modulation of the current and
the surface elevation also introduce an additional scattering, via the $M_c$
term in the coupling coefficent (\ref{M}).
 Notations here assume that $\kb$ is in the direction of the current and $\kb'$
 is opposite to the current.
At resonance, in the limit $\eta\rightarrow 0$, the amplitude reflection
coefficient $R_S$ given by (\ref{Krsbscat}) converges to the reflection
coefficient given by Kirby (1988). Using our notations, he obtained
\begin{equation}
R_{\rm Kirby}
=\left[\frac{\sigma'\left(Cg+U\right)}{\sigma\left(Cg'-U\right)}\right]^{1/2}\tanh(QL)\label{KKirby},
\end{equation}
with
\begin{equation}
Q=\frac{\Omega_c \omega}{\left[\sigma
\sigma'\left(Cg+U\right)\left(Cg'-U\right) \right]^{1/2}}
\end{equation}
and $\Omega_c = -M(k,k') b /\left[4  \omega F^B(k-k') \right]$. Our amplitude
reflection coefficient $R_{S}$ is estimated with the approximation $\er^{q
\sfbsQ L} = \left(\sfbsI+q \sfbsQ\right) L + O\left((qL)^2\right)$, so that, to
first order in $qL$,
\begin{equation}
R_{S} \approx \left[\frac{\sigma' C_g C_g' q L }{\sigma} \right]^{1/2}.
\end{equation}
 Replacing the
analytical expression (\ref{specatres}) in (\ref{defq}) yields
\begin{equation}
R_{S} \approx \frac{bLM(k,k')}{4 \left[\sigma^2 (C_g'-U)^2 \right]^{1/2}},
\end{equation}
which is clearly identical to(\ref{KKirby}) at first order in $qL$.

For finite values of $qL$, the reflection coefficient (\ref{Krsbscat})
corresponding to the solution of (\ref{Nevol_mono}) is obtained by calculating
the proper matrix exponential. Anticipating oceanographic conditions with a
water depth of 20 m, a strong 2 m~s$^{-1}$ current corresponds to a Froude
number of 0.17 only. For such a low value of $\Frou$ in the context of  Davies
\& Heathershaw's (1984) laboratory experiments, the convergence of the present
theory and that of Kirby (1988) is illustrated in figure 5.  The reflection
coefficient is largely increased for following currents due to the general
conservation of the wave action flux. In that case $R$ is enhanced by the
factor $\left\{\sigma (Cg+U) / \left[\sigma'(Cg'-U)\right]\right\}^{1/2}$. The
overall increase in $R$ for following waves amounts to about 60\% at
$\Frou=0.17$, for the laboratory sinusoidal bars of Davies \& Heathershaw
(1984) shown before (figure 3), with a reflected wave energy multiplied by a
factor 2.5, compared to the case without current.
%%%%%%%%%%%%%%%%%%%%%%%%%%%%%%%%%%%%%%%%%%%%%%%%%%%%%%%%%%%%%%%%%%%%%%%%%%%%%%%%
\begin{figure}
%\epsfig{file=bottomdef.eps,width=\linewidth}
%\centerline{\includegraphics[width=\textwidth]{bottomdef.eps}}
\centerline{\includegraphics[width=0.9\textwidth]{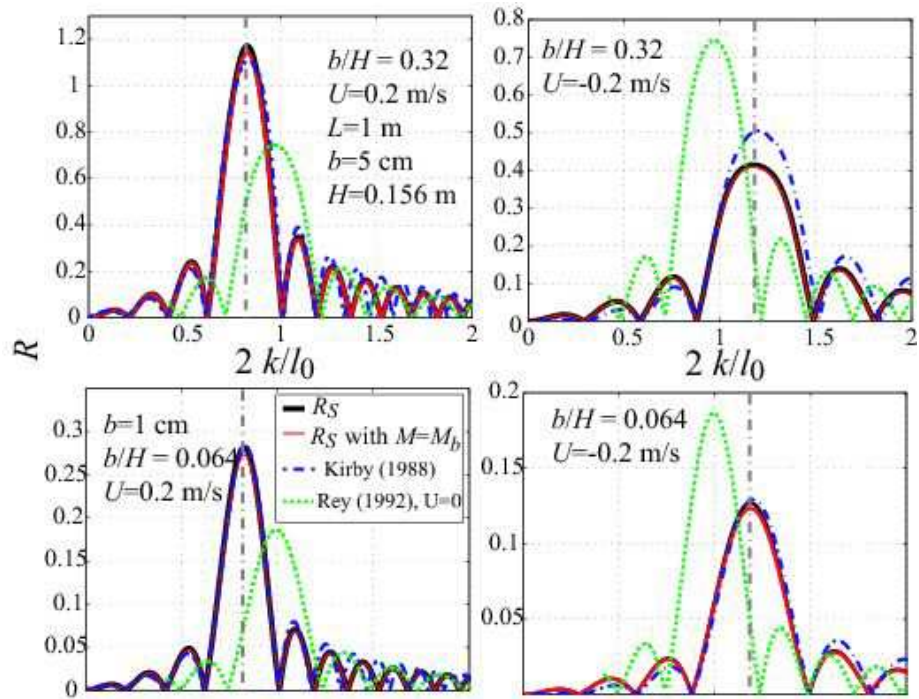}}
\caption{Amplitude reflection coefficients for monochromatic waves over
sinusoidal bars for the same settings as in figure 3, with a following (left)
or opposing (right) current of magnitude $U=0.2$~m~s$^{-1}$. For reference the
reflection coefficient without current, as given  by the exact model of Rey
(1995), is also shown. The position of the resonant wavenumber is indicated
with the grey vertical dash-dotted line.} \label{figKirby}
\end{figure}
%%%%%%%%%%%%%%%%%%%%%%%%%%%%%%%%%%%%%%%%%%%%%%%%%%%%%%%%%%%%%%%%%%%%%%%%%%%%%%%%
For this mild current the contribution of the current fluctuation to the
coupling coefficient is small, with a maximum increase of 16\% on the action
reflection coefficent, 8\% for the wave amplitude. However, for larger Froude
numbers, this additional scattering may become significant as illustrated by
figure \ref{figKirby2}. The present theory and that of Kirby (1988) agree
reasonably well for finite values of $\eta$, and we thus expect the source term
to represent accurately the scattering of waves over bottom topographies in
cases of uniform currents.

For $m=4$ sinusoidal bars, the energy reflection coefficients was found to be
within 10\% of the exact solution for over 90\% of the wavenumber range shown
in figure 3, for $\eta<0.1$ and $\Frou =0$, and this conclusion is expected to
hold for $\Frou < 0.2$, given the agreement with Kirby's (1988) approximate
solution. This accuracy is twice better than what was found for a rectangular
step with $\Frou = 0$ (MAHR). The present method has the advantage of a large
economy in computing power. This method is also well adapted for natural sea
beds, for which continuous bathymetric coverage is only available in restricted
areas, and thus only the statistical properties of the bottom topography are
accessible, assuming homogeneity.

%%%%%%%%%%%%%%%%%%%%%%%%%%%%%%%%%%%%%%%%%%%%%%%%%%%%%%%%%%%%%%%%%%%%%%%%%%%%%%%%
\begin{figure}
%\epsfig{file=bottomdef.eps,width=\linewidth}
%\centerline{\includegraphics[width=\textwidth]{bottomdef.eps}}
\centerline{\includegraphics[width=0.7\textwidth]{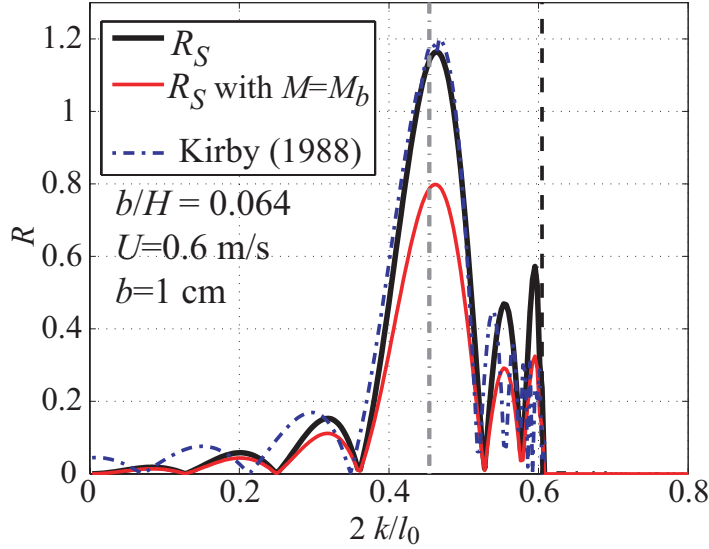}}
\caption{Amplitude reflection coefficients for monochromatic waves over
sinusoidal bars for the same settings as in figure 3 and 4, with a stronger
following current of magnitude $U=0.6$~m~s$^{-1}$. The position of the resonant
wavenumber is indicated with the grey vertical dash-dotted line. The vertical
dashed line corresponds to the wavenumber for which $Cg'=U$. For larger
wavenumbers the  reflected waves are blocked and cannot propagate against the
current.} \label{figKirby2}
\end{figure}
%%%%%%%%%%%%%%%%%%%%%%%%%%%%%%%%%%%%%%%%%%%%%%%%%%%%%%%%%%%%%%%%%%%%%%%%%%%%%%%%

\section{Scattering with current on a realistic topography}
\subsection{Sandwaves in the North Sea}
\begin{figure}
%\epsfig{file=bottomdef.eps,width=\linewidth}
%\centerline{\includegraphics[width=\textwidth]{bottomdef.eps}}
\centerline{\includegraphics[width=\textwidth]{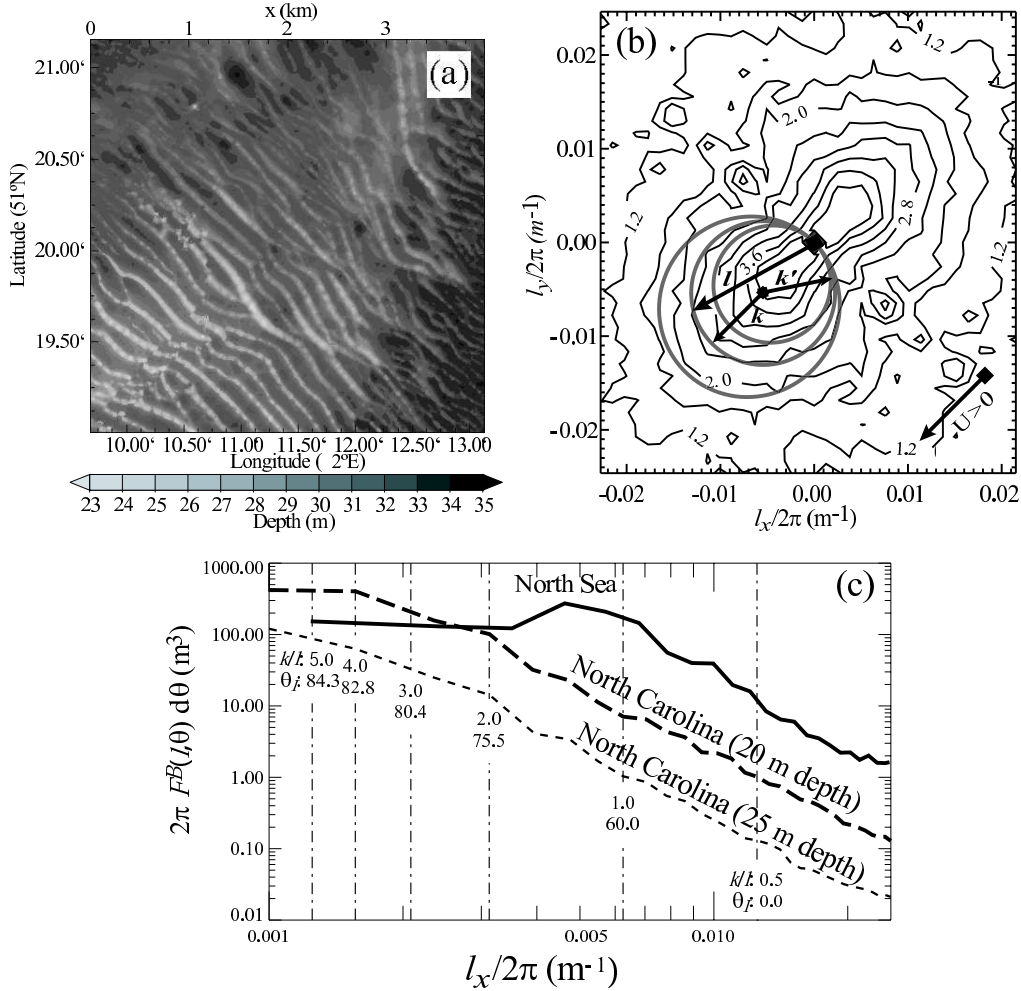}}
\caption{(a) high-resolution bathymetry of a sand wave field in the southern
North Sea with depths relative to chart datum, and (b) corresponding bottom
elevation spectrum with contour values representing $\log_{10}\left( 4\upi
^{2}F^{B}\right)$. The locus of the interacting bottom and surface wave
components are indicated for a 12.5~s waves from the North-East in 25 m depth,
with $U=0$ (middle circle), $U=2$~m~s$^{-1}$ (smaller ellipse), and
$U=-2$~m~s$^{-1}$ (larger ellipse), $U$ is positive from the North-East. (c)
Direction-integrated bottom variance spectra from the North Carolina shelf and
the southern North Sea. Vertical lines indicate $k/l$ ratios and incident
resonant directions $\theta_I$, assuming an incident wave field of 12.5~s
period in 25~m depth and bedforms parallel to the $y$-axis. For such bedforms,
the angle between incident and scattered waves is $180^\circ-2\theta_I$.}
\label{bspectra}
\end{figure}

A real ocean topography, at least on the continental shelf, generally presents
a continuous and broad bottom elevation spectrum. The effects of a mean current
on wave scattering are now examined using a bottom spectrum estimated from a
detailed bathymetric survey of an area centered on the crest of a sand dune, in
the southern North Sea (figure \ref{bspectra}). In this region, tidal currents
are known to generate a wide array of bedforms, from large scale tidal Banks to
sand dunes and sand waves (e.g. Dyer \& Huntley 1999; Hulscher \& van den Brink
2001\nocite{Dyer&Huntley1999,Hulscher&vandenBrink2001}). Although sand dunes
present a threat to navigation and are closely monitored (Idier \etal~2002),
dunes are much larger than typical wind sea and swell wavelengths. These dunes,
however, are generally covered with shorter sandwaves. In the surveyed area the
sandwaves have a peak wavelength of 250~m, and an elevation variance of
1.7~m$^2$, which should lead to strong oblique scattering of waves with periods
of 10~s and longer. Over smaller areas of 3 by 3 km the variance can be as
large as 3.3~m$^2$ with a better defined spectral peak, so that our chosen
spectrum is expected to be representative of the entire region, including high
and low variances on dunes crests and troughs, respectively.  The southern
North Sea is also known for the attenuation of long swells, generated in the
Norwegian Sea. This attenuation has been generally attributed to the
dissipation of wave energy by bottom friction (Weber 1991\nocite{Weber1991b}).

The bottom spectrum of the chosen area, like the spectra that were obtained by
AH from the North Carolina shelf, rolls off sharply at high wavenumbers,
typically like $l^{-3}$ for the directionally-integrated bottom spectrum
$F^{B2D}$, and proportional to $l^{-4}$ for the full spectrum $F^B$. Here the
maximum variance is found for bottom wavelengths of the order of or larger than
250~m (figure \ref{bspectra}). For a typical swell period of 10~s, this
corresponds to 2 times the wavelength in 20~m depth, and thus a rather small
scattering angle, 30$^\circ$ off from the incident direction. Swells
propagating from a distant storm, with fixed absolute frequency $\omega=\sigma
+ \kb \bcdot \Ub$, should be reflected by bottom undulations with widely
different variances as the current changes.
%%%%%%%%%%%%%%%%%%%%%%%%%%%%%%%%%%%%%%%%%%%%%%%%%%%%%%%%%%%%%%%%%%%%%%%%%%%%%%%%
%%%%%%%%%%%%%%%%%%%%%%%%%%%%%%%%%%%%%%%%%%%%%%%%%%%%%%%%%%%%%%%%%%%%%%%%%%%%%%%%

Given this bottom spectrum and the mean water depth, simple solutions are
available for uniform conditions, because the scattering source term is a
linear function of the directional spectrum at a given value of the absolute
frequency $\omega$ (see AH for numerical methods). We consider the wave
directional spectrum for a frequency $f_0$ and discretize it in $N_a$
directions. This spectrum is thus a vector ${\mathbf E}$ in a space with $N_a$
dimensions. The square matrix $ {\sfbsS}$ such that ${\rm d} {\mathbf E}/{\rm
d} t = {\sfbsS} {\mathbf E}$ is symmetric and positive, and can thus be
diagonalized, which gives $N_a$ eigenvalues $\lambda_n$ and corresponding
eigenvectors ${\mathbf V}_n$, such that ${\sfbsS} {\mathbf V}_n = \lambda_n
{\mathbf V}_n$. Thus the time evolution is easily obtained by a projection of
${\mathbf E}$ on the basis $\{{\mathbf V}_n, 1 \leq n \leq N_a \}$, giving a
decomposition of ${\mathbf E}$ in elementary components. Each of these
components of the directional spectrum decays exponentially in time, except for
the isotropic part of the spectrum which remains constant because that
eigenvector corresponds to $\lambda =0$. The eigenvalues thus give interesting
timescales for the evolution of the spectrum toward this isotropic state, with
a half-life time of each eigenvector given by $-\ln 2/\lambda_n$.

Numerical results are shown here for a mean water depth of 20~m, in order to
make the result more visible. For that depth, waves with a period $T=10$~s have
a dimensionless depth $kH=1.04$, which is close the value for which the
coupling coefficient $M_b$ is maximum (AH). As a result, scattering is probably
stronger than in real conditions where the mean water depth is 30~m. The
following results should still provide some understanding of the likely real
effects, at least for larger wave periods with similar values of $kH$. Without
current, if $kH$ is kept constant, the magnitude of the coupling coefficient
$K(k,k',H)$ decreases like $H^{-9/2}$ (AH), but it is compounded by a higher
bottom elevation spectral density for small values of $k$. For back-scattering,
the bottom wavenumbers are generally in the range where the bottom spectrum
rolls off like $l^{-4}$ (figure \ref{bspectra}). Therefore, for these
back-scattering directions, the evolution time scale of waves with the same
value of $kH$, e.g. $T=11.2$~s in 25~m depth or $T=13.2$~s in 35~m depth,  is
larger by a factor $(25/20)^{1/2}\simeq 1.1$ or $(35/20)^{1/2}\simeq 1.3$,
respectively. For incident wave and scattering directions for which the bottom
spectrum is more uniform and does not compensate for the reduction in the
coupling coefficient, such as forward scattering of waves from the North-West,
the time scales increase by $(25/20)^{9/2}\simeq 3.7$ or $(35/20)^{9/2}\simeq
77$, respectively.

With $N_a=120$, corresponding to a directional resolution of $3^\circ$, figure
8 shows that the shortest time scales (large negative values of $\lambda_n$)
correspond to directional spectra (eigenvectors) with strong local variations.
These eigenvectors are thus associated with scattering at small oblique angles
(forward scattering). Only the last 10 eigenvalues have a rather broad support,
corresponding to scattering at much larger angles. Besides, the strongest
scattering corresponds to a half-life time of 430~s, and mostly affects waves
from the North-West or South-East, i.e. propagating in a direction along the
sandwave crests. The timescale for waves from the North-East or South-West is
about five times larger (the corresponding range of indices is $80<n<110$).
%%%%%%%%%%%%%%%%%%%%%%%%%%%%%%%%%%%%%%%%%%%%%%%%%%%%%%%%%%%%%%%%%%%%%%%%%%%%%%%%
\begin{figure}
%\epsfig{file=bottomdef.eps,width=\linewidth}
%\centerline{\includegraphics[width=\textwidth]{bottomdef.eps}}
\centerline{\includegraphics[width=0.8\textwidth]{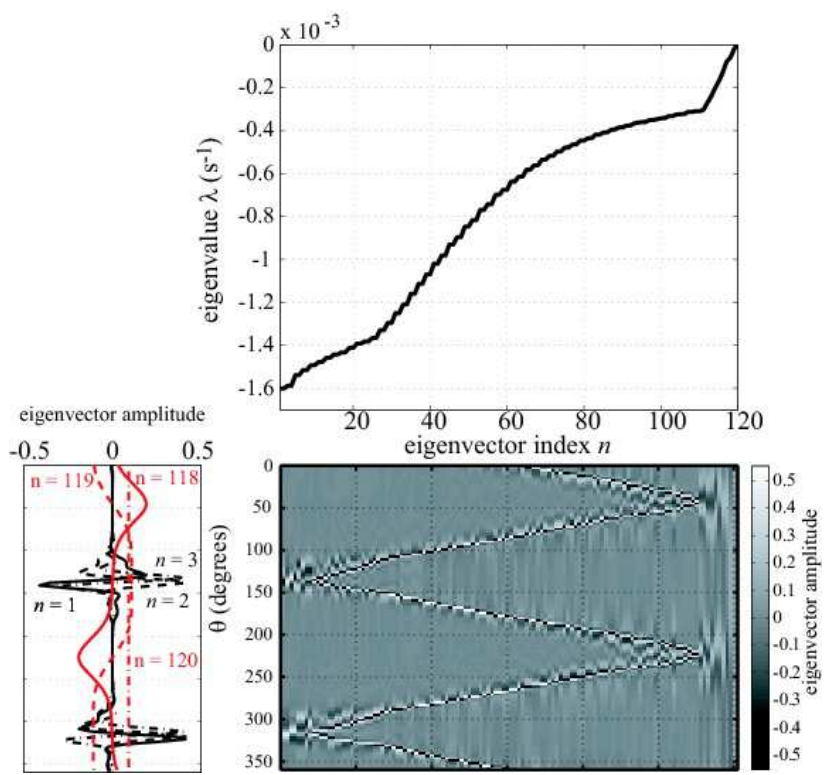}}
\caption{Eigenvalues ordered by magnitude (top), and corresponding eigenvectors
(bottom right) of the scattering matrix $ {\sfbsS}$ for $U=0$, $f_0=0.1$~Hz,
and $H=20$~m. The first three and last three eigenvectors are shown in more
detail in the bottom left.} \label{figEigen}
\end{figure}
%%%%%%%%%%%%%%%%%%%%%%%%%%%%%%%%%%%%%%%%%%%%%%%%%%%%%%%%%%%%%%%%%%%%%%%%%%%%%%%%
The $n=118$ eigenvector corresponds to an exchange of wave energy between waves
travelling in opposite directions across the sandwaves, but the corresponding
half-life is of 3 hours and 15 minutes. Similar results were found for
$N_a=180$ and $N_a=72$ and appear little sensitive to the discretization.

Instead of this idealized horizontally uniform situation, practical situations
rather correspond to quasi-stationary conditions with spatial gradients in at
least one dimension. In this case the simple steady solutions found above for
2D topography are not physical. Indeed, a 3D bottom causes scattering along the
transversal direction $y$, and the energy propagating in that direction builds
up slowly up to the point where it becomes as large as the incident wave
energy. This process can take a time much longer than the typical duration of a
storm or swell arrival, and dissipative processes are likely to be important as
the wave energy increases (e.g. Ardhuin \etal~2003). In order to go beyond
qualitative statements on time and spatial scales of spectral relaxation, and
short of simulating an actual storm in two dimensions, the effects on the wave
spectrum are illustrated with a one-dimensional model configuration.

The source term $S_{\mathrm{bscat}}$ was introduced in the version $2.22$ of
the wave model WAVEWATCH~III (Tolman 1991,
2002\nocite{Tolman1991b,Tolman2002c}), based on the wave action evolution
equation (\ref{action_balance}) in which the time derivative on the left hand
side is now a Lagrangian derivative following a wave packet in physical and
spectral space. Bottom scattering is the only source term activated in the
present calculation.
 The
 model was run with a spectral grid of $30$ frequencies ranging from $0.04$ to $0.788$~Hz
 and a directional resolution of $3^\circ$.
 Unfortunately the model spectrum
 is discretized  with components at fixed intrinsic frequencies $\sigma$ and directions
$\theta$, which is most appropriate for other processes. Therefore a small
amount of numerical diffusion leads to a change of action at each absolute
frequencies when $U\neq 0$, and the total action is only approximately
conserved in that case, with a net change of about $1\%$ of the integral of the
absolute value of the source term for $U=\pm 2$~m~s$^{-1}$, and four orders of
magnitudes smaller, i.e. at the round-off error level, for $U=0$. We have
chosen to show cases with
 significant back-scatter, corresponding to waves normally incident over
 the sandwaves. This choice also corresponds to a weaker forward scattering,
 compared to waves propagating along the the sandwave crests.

\subsection{Scattering of waves normally incident on the sandwaves}
\begin{figure}
%\epsfig{file=bottomdef.eps,width=\linewidth}
%\centerline{\includegraphics[width=\textwidth]{bottomdef.eps}}
\centerline{\includegraphics[width=\textwidth]{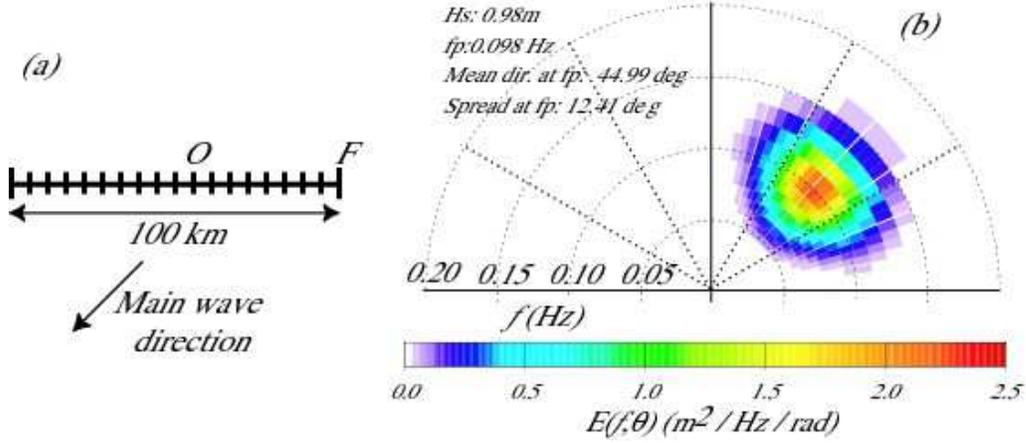}} \caption{(a)
Schematic of the model grid and  (b) incident wave spectrum specified at point
$F$. Model output is shown below for point $O$.  Please note that waves are
represented with their arrival direction (direction from, contrary to the
standard wind sea convention). The frequency is the relative frequency $\sigma
/ 2\upi$.} \label{incidentwave}
\end{figure}

To simplify the interpretation of the results, and the processing of the
boundary conditions, a one dimensional (East-West) propagation grid is used for
the computations, assuming that the wave field, still fully directional, is
uniform in the North-South direction. The waves are propagated over a model
grid $100$~km long, with a mean depth of $H=20$m, and a spatial grid step of
5~km (figure \ref{incidentwave}.a). As discussed above, this water depth is
chosen to make the result more visible, and a significant broadening of the
incident peak with a (weaker) back-scatter of waves is also found for $H=35$~m
and $f_p=0.1$~Hz (not shown).
%%%%%%%%%%%%%%%%%%%%%%%%%%%%%%%%%%%%%%%%%%%%%%%%%%%%%%%%%%%%%%%%%%%%%%%%%%%%%%%%
%%%%%%%%%%%%%%%%%%%%%%%%%%%%%%%%%%%%%%%%%%%%%%%%%%%%%%%%%%%%%%%%%%%%%%%%%%%%%%%%
A Gaussian incident surface
 wave spectrum is imposed, with a mean direction from the North-East, a narrow peak directional spread of
 $12^\circ$, and a peak frequency of $0.01$~Hz (figure \ref{incidentwave}.b).  The source term is integrated
 with a time step
 of $120$~s, and the
 advection in space uses a third order scheme with a time step of
 $120$~s (Tolman 2002).

The scattering source term acts as a diffusion operator with a typical 3-lobe
structure, negative at the peak of the wave spectrum, and positive in
directions of about 30$^\circ$ on both sides of the peak. This is identical,
but with a larger magnitude, to the effect described by AH. In general the
scattering effects are relatively stronger at the lowest frequencies, at least
in the range of frequencies used here. For still lower frequencies the
scattering coefficient $K$ decreases (see also AH) so that, on these spatial
scales, very little scattering occurs for infra-gravity waves ($f< 0.05$~Hz).
In addition to this grazing-angle forward scattering, a significant
back-scatter is found, in particular in the case of following currents.
%%%%%%%%%%%%%%%%%%%%%%%%%%%%%%%%%%%%%%%%%%%%%%%%%%%%%%%%%%%%%%%%%%%%%%%%%%%%%%%%
\begin{figure}
%\epsfig{file=bottomdef.eps,width=\linewidth}
%\centerline{\includegraphics[width=\textwidth]{bottomdef.eps}}
\centerline{\includegraphics[width=0.8\textwidth]{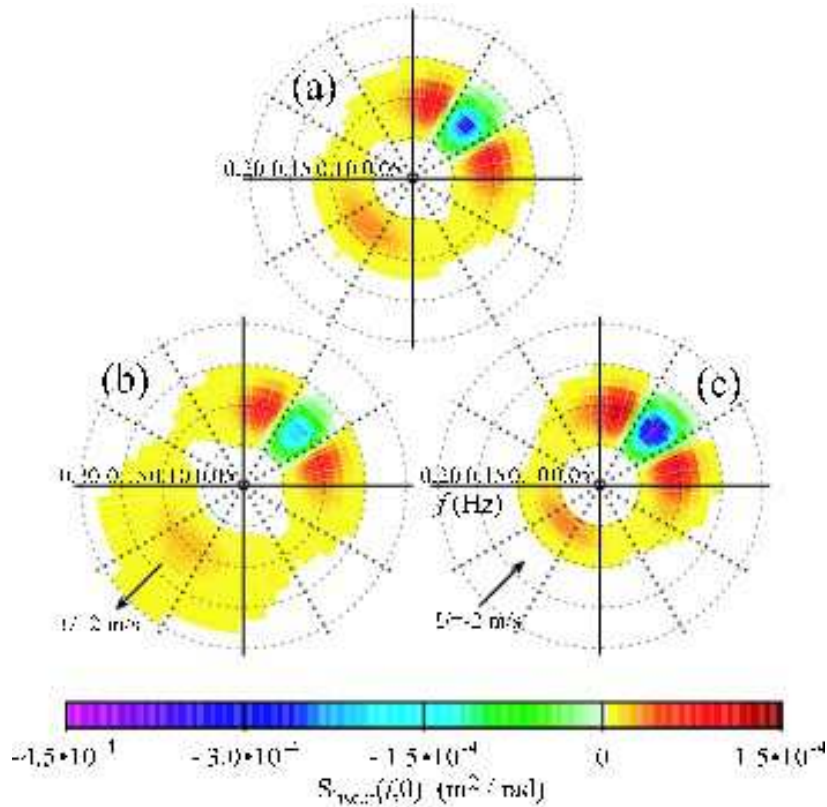}}
\caption{Computed source terms at the boundary forcing point $F$, (a) for
$U=0$, (b) for a following current $U=2$~m~s$^{-1}$, (c) for an opposing
current $U=-2$~m~s$^{-1}$. The frequency is the relative frequency $f=\sigma /
2\upi$.}\label{source}
\end{figure}
%%%%%%%%%%%%%%%%%%%%%%%%%%%%%%%%%%%%%%%%%%%%%%%%%%%%%%%%%%%%%%%%%%%%%%%%%%%%%%%%

%%%%%%%%%%%%%%%%%%%%%%%%%%%%%%%%%%%%%%%%%%%%%%%%%%%%%%%%%%%%%%%%%%%%%%%%%%%%%%%%
\begin{figure}
%\epsfig{file=bottomdef.eps,width=\linewidth}
%\centerline{\includegraphics[width=\textwidth]{bottomdef.eps}}
\centerline{\includegraphics[width=0.8\textwidth]{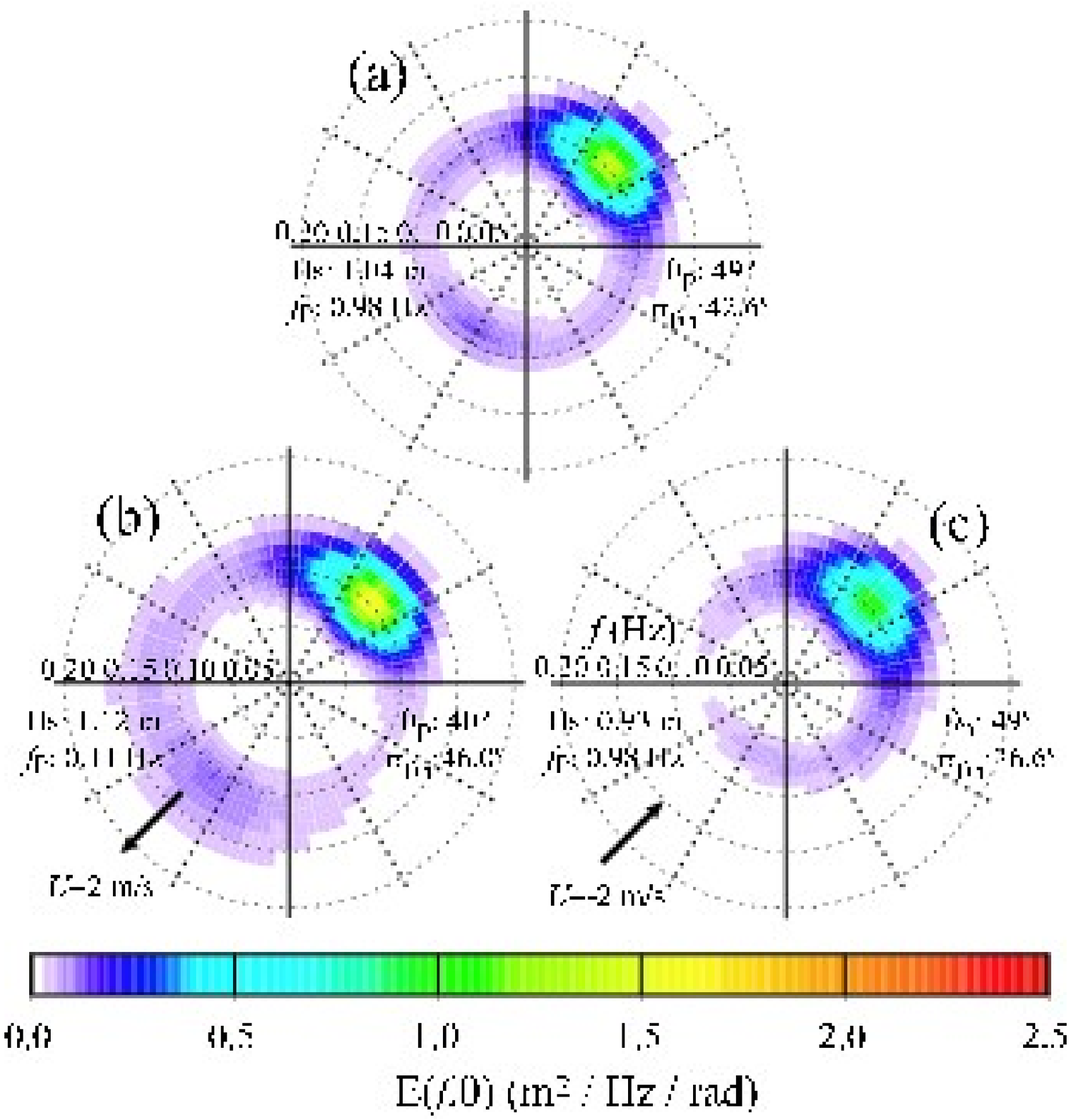}}
\caption{Computed wave spectra at point $O$, 40~km inside of the model domain,
after 5~hours of propagation, (a) for $U=0$, (b) for a following current
$U=2$~m~s$^{-1}$, (c) for an opposing current $U=-2$~m~s$^{-1}$. The frequency
is the relative frequency $f=\sigma / 2\upi$.}\label{spectra}
\end{figure}
%%%%%%%%%%%%%%%%%%%%%%%%%%%%%%%%%%%%%%%%%%%%%%%%%%%%%%%%%%%%%%%%%%%%%%%%%%%%%%%%

 For an absolute wave frequency of $0.08$~Hz, the curves followed by the
bottom resonant wavenumbers are overlaid on the bottom spectrum (figure
\ref{bspectra}.b). The wavenumbers
 $\lb$ along these curves satisfy both
the relations $\kpb+\lb=\kb$ and $\sigma'=\sigma+\lb \bcdot \Ub$. Without
current the curve is exactly a circle, and transforms to an ellipse for
relatively weak currents (Appendix C). This approximation is used in the model
to compute the source term. The current imposed here shifts significantly the
resonant configuration for the bottom and surface wavenumbers. A current
opposed to the waves enlarges the ellipse towards higher wavenumbers, while a
following current will lead to a `sampling' of shorter wave numbers, i.e.
bottom features of larger scales. Since the bottom topography has the largest
variance at low wavenumbers, scattering is strongest for following currents
(figure \ref{source}). With our choice of parameters, there is about a factor
10 reduction in the bottom variance that causes backscatter as $U$ is changed
from $2$~m~s$^{-1}$ to $-2$~m~s$^{-1}$. Besides, the coupling coefficient
$K(k',k,H)$ is increased in the case of a following current, as discussed above
for the 2D cases.

The resulting wave spectra are also modified due to the conservation of the
wave action flux, enhancing the reflected wave energies for $U>0$ (figure
\ref{spectra}). This effect is similar to what was found in the 2D cases
considered above, due to the different energy flux velocities $U+C_g$ for the
incident waves, and $U-C_g'$ for the reflected waves. In all cases investigated
here, the narrow incident wave spectrum is significantly broadened in
directions, and that effect is most pronounced for frequencies in the range
0.07--0.10~Hz. Without current or with following currents, spectra in the
middle of the model domain exhibit a significant level of back-scattered
energy, which increases the significant wave height and the directional spread
on the up-wave side of the sandwave field (figure \ref{spectra}). This effect
should not be very sensitive to the directional spread of the incident wave
field, because the projection of the directional spectrum on the corresponding
`smooth' eigenvectors of the scattering matrix (figure 8) is insensitive to
local variations in the directional spectrum. This reflection should thus occur
for a wide range of sea states. At the same time, the incident peak of the wave
field broadens in directions as it propagates to the down-wave end of the model
domain. This broadening is fast close the the forcing boundary (point F), with
values of the peak frequency directional spreads $\sigma_{\theta,p}$ larger
than $35^\circ$ at a point 5~km inside the domain (not shown), and becomes more
gradual as the waves propagate, due to the slower evolution of broad spectra
that are associated with smaller eigenvalues in the scattering matrix (see also
Ardhuin \etal~2003a, Ardhuin \& Herbers 2005\nocite{Ardhuin&Herbers2005}). It
was also verified that this broadening of the main spectral peak is strongest
for waves propagating along the main sandwave crest directions (e.g. from the
North-West in our case) due to the larger bottom variance at $\lb=\kb-\kb'$
with $\kb\simeq\kb'$, resulting in a significant modification of the mean
direction (Magne 2005\nocite{Magne2005}).

Finally, a decrease in significant wave height is found along the grid,
indicating an attenuation due to wave-bottom scattering. In reality, bottom
friction would likely induce a stronger decay, and that decay would be stronger
than in the absence of scattering. Essentially the scattering increases the
average time taken by wave energy to cross the domain, and, because of that
longer time, bottom friction together with scattering would lead to a larger
dissipation than friction alone (Ardhuin \etal~2003).

\section{Conclusion}
The effect of a uniform current on the scattering of random surface gravity
waves was investigated theoretically, extending the derivations of Ardhuin \&
Herbers (2002). Wave scattering may thus be represented by a scattering source
term $S_{\rm bscat}(\kb)$ for each wave component $\kb$, in a closed spectral
action balance equation. That term gives the rate of exchange of wave action
between wave components $\kb$ and $\kpb$ that have the same absolute frequency,
as a result of both water depth variations on the scale of the surface gravity
waves wavelength, and current and mean free surface inhomogeneities induced by
the bottom topography. The exchange of action between any two wave component
pairs $\kb$ and $\kpb$ is proportional to the bottom elevation spectrum at the
wavenumber vector $\lb=\kb-\kpb$, which is characteristic of Bragg scattering.
The spectral integral of the corresponding wave pseudo-momentum source term
$\kb S_{\rm bscat}$ gives a recoil force exerted by the bottom on the water
column, in addition to the hydrostatic pressure force.

After Magne \etal~(2005a) proved that the source term was applicable to
non-random topography and accurate in the limit of small bottom amplitudes,
just like Bragg scattering approximations for acoustic or electromagnetic waves
(e.g. Elfouhaily \& Guerin 2004), it is further found here that monochromatic
wave results are recovered by taking the limit to narrow incident and reflected
wave spectra. In absence of current, for a finite sinusoidal bottom and
monochromatic waves, the reflection coefficients given by the source term
converges to Mei's (1985) theory in the limit of the small bottom amplitudes.
The range of maximum reflection and the side lobe pattern of the reflection
coefficient as a function of the incident wavenumber is thus a direct
consequence of the shape of the bottom spectrum in that case. With this point
of view, there is resonance at all wavenumbers but its strength is proportional
to the bottom elevation variance at the corresponding scale. In the presence of
a current, reflections converge in the same manner to the more general theory
of Kirby (1988). In two dimensions, the main effects of a current is an
enhancement of reflected wave amplitudes when the incident waves propagate with
the current, due to a conservation of the wave action flux, and a Doppler-like
shift of the resonant wave frequencies that undergo maximum reflection. The two
scale approximation was found to hold very well, even for a relatively fast
evolutions of the wave amplitudes over two wavelengths (e.g. figure 3).
However, the source term does not give a good representation of the spatial
evolution of the wave field on scales shorter that the bottom correlation
length, nor can it give reasonable results when another wave train propagates
from beyond the bars. In that latter case, a lower order source term must be
considered, and a closed action balance cannot be obtained since that extra
term depends on the phase relationship between the incident waves, reflected
waves and bottom undulations.

 In three dimension and over the shallow areas of the southern North Sea,
where large sand waves are found with strong tidal currents, wave scattering is
expected to be significant, and largely influenced by currents. Over natural
topographies, the bottom typically de-correlates over scales shorter than the
scattering-induced attenuation scales, so that a modification of the reflection
due to a phase locking of the incident and reflected waves with the bottom can
be neglected. The wave scattering theory presented in this paper is thus one
more piece in the puzzle of wave propagation over shallow continental shelves,
and this process may account for a significant part of the observed attenuation
of swells in the southern North Sea. The representation of this phenomenon with
a source term in the wave action balance equation is expected to be accurate in
many conditions of interest. It is consistent with the wide use of
phase-averaged models for engineering and scientific purposes when such large
scales are involved. The alternative use of phase-resolving elliptic
refraction-diffraction models (e.g. Belibassakis \etal
2001\nocite{Belibassakis&al.2001}), is much more expensive in terms of computer
resources,  due to the necessity to resolve the wave phase and the ellipticity
of the problem when back-scattering occurs. %Fu for a depth-varying current, $U$ should
%be regarded as the wave advection velocity (Andrews \& McIntyre
%1978\nocite{Andrews&McIntyre1978b}, see Kirby \& Chen
%1989\nocite{Kirby&Chen1989} for practical approximate expressions).
For applications to rotational currents, the mean current $U$ should be
regarded as the wave advection velocity (Andrews \& McIntyre
1978\nocite{Andrews&McIntyre1978b}, see Kirby \& Chen
1989\nocite{Kirby&Chen1989} for practical approximate expressions), but a
detailed derivation including scattering by rotational current fluctuations
should be the next logical extension of the present theory. This is probably
achievable by coupling the rotational part of the flow to the irrotational
part, giving a modified Bernoulli equation (e.g. McWilliams
\etal~2004\nocite{McWilliams&al.2004}). In practice, non-homogeneities in the
bottom spectrum will probably have to be addressed due the sharp decrease of
the coupling coefficient with water depth, and the generally higher bottom
elevation variances in the shallower parts of the sea floor. In particular our
limited bathymetric survey shows that sandwaves are modulated by sand dunes,
very much like short water waves are modulated by long waves.

\begin{acknowledgments}
This research was supported by a joint grant from CNRS and DGA. Bathymetric
data was acquired by the French Hydrographic and Oceanographic Service (SHOM).
Discussions with Michael McIntyre, Kostas Belibassakis, Vincent Rey,  and
Thierry Garlan and gratefully acknowledged. The results of the relative effects
of current modulations and water depths changes owes much to remarks made by
anonymous reviewers, without whom the present paper would have been limited to
small Froude numbers.
\end{acknowledgments}

 \appendix
 \section{Harmonic oscillator equation for the first order potential}
 The harmonic oscillator equation (\ref{oscil2}) can be written as a
 linear superposition of equations of the type
 \begin{equation}\label{A1}
\frac{d^2 f_1}{dt^2}+\omega^2 f_1=\er^{\ir \omega't}.
 \end{equation}
In order to specify a unique solution to (\ref{A1}), initial conditions must be
prescribed. In the limit of the large propagations distances, the initial
conditions contribute a negligible non-secular term to the solution. Following
Hasselmann (1962)\nocite{Hasselmann1962}, we choose $f_1(0)=0$  and
$df_1/dt(0)=0$, giving,
\begin{equation}
  f_1(\omega,\omega'; t)= \frac{\er^{{\mathrm i}\omega' t}-\er^{{\mathrm i}\omega t}+i(\omega-\omega')\sin(\omega t)/\omega}{\omega^2-\omega'^2}
  \mbox{   for }\omega'^2\neq\omega^2,
\end{equation}
\begin{equation}
f_1(\omega,\omega'; t)= \frac{t\er^{{\mathrm
i}\omega't}}{2i\omega'}-\frac{\sin'\omega t)}{2i\omega'\omega} \mbox{   for
}\omega'=\pm\omega
\end{equation}

 \section{Harmonic oscillator equation and energy for the second order potential}
Replacing $\phi_1$ (\ref{formphi2}) in the surface boundary condition
(\ref{surfbound3}),
\begin{equation}
\left( \frac{ d^2}{dt^2}+\sigma^2\right) \Phi^s_{2,\kb}(t)= -gk\Phi^{{\rm
si},s}_{2,\kb} -\tanh(kH) \frac{\partial^2\Phi^{{\rm si},s}_{2,\kb}}{\partial
t^2}+{\rm I-VIII},
\end{equation}
and conserving only the resonant terms of $\Phi_{1,\kb'}^s$, one obtains
 \begin{eqnarray}
 \frac{\partial^2\Phi^{{\rm si},s}_{2,\kb}}{\partial t^2}=&& \nonumber\\
 -\sum_{\kpb,\kb''} \frac{\kpb \cdot
 \kb}{k}& &\frac{\cosh(kH)}{\cosh(k'H)}
 M(\kpb,\kb'')G_{\kb-\kpb}G_{\kpb-\kb''}
 \Phi_{0,\kb''}\frac{\partial^2}{\partial t^2}
 \left(f_1(\sigma',\lb' \bcdot \Ub-s\sigma'')\er^{\ir \lb
\bcdot \Ub t} \right), \nonumber\\
 \end{eqnarray}
with $\lb'=(\kb''-\kpb)\bcdot \Ub$. In order to simplify the algebra we assume
that the zeroth-order waves are random, with no correlation between
$\Phi_{0,\kb}^s$ and $\Phi_{0,\kb''}^{s''}$ unless $\kb=\pm \kb''$ and $s=\pm
s''$. Thus the only contributing terms to $N_{2,0}$ must verify $\kb''=\kb$.
Only those terms are now written explicitly, the others being grouped in the
'$\ldots$'. The amplitude $\Phi^+_{2,\kb}$ satisfies the following forced
harmonic oscillator equation,
\begin{eqnarray}\label{dd1}
\left( \frac{\partial^2}{\partial t^2}+  \sigma^2\right) \Phi^+_{2,\kb}(t)
=\sum_{\kpb} M^2(\kb,\kpb) \left|G_{\kb-\kpb}\right|^2
\Phi_{1,\kb''}f_1(\sigma',-\sigma-\lb \bcdot \Ub ) \er^{\ir
\lb \bcdot \Ub t}+\ldots \nonumber\\
\end{eqnarray}
This is a sum of equations of the form,
\begin{equation}\label{f2}
\left( \frac{ d^2}{dt^2}+\sigma^2\right)f_2=f_1(\sigma',\lb \bcdot \Ub
-\sigma;t)  \er^{\ir \lb \bcdot \Ub t}.
\end{equation}
The solution $f_2$ may be written as
 \begin{equation}
   f_2=f_{2,a}+f_{2,b},
 \end{equation}
 where
 \begin{equation}\label{f2a}
   f_{2,a}=-\frac{t\er^{-\ir \sigma t}-\sin(\sigma
   t)/\sigma}{2 \ir \sigma \left[\sigma'^2-(\lb \bcdot \Ub+\sigma)^2\right]},
 \end{equation}

 \begin{eqnarray}\label{f2b}
   f_{2,b}& =&-\frac{1}{2\sigma'\left[\sigma'-(\lb \bcdot \Ub+\sigma )\right]} \times \nonumber\\
   &  & \left[ \frac{\er^{-\ir (\sigma'-\lb \bcdot \Ub)t}}{\sigma^2-(\sigma'-\lb \bcdot \Ub)^2}
  -\frac{1}{2\sigma}
  \left( \frac{\er^{{\mathrm i}\sigma t}}{\sigma+(\sigma'-\lb \bcdot \Ub)}+\frac{\er^{-\ir \sigma t}}
  {\sigma-(\sigma'-\lb \bcdot \Ub)} \right)
   \right]
 \end{eqnarray}

 The second order action contribution from correlation between the zeroth and
first order velocity potential is given by,
\begin{equation}
  F^{\Phi}_{2,0,\kb}= F^{\Phi}_{0,2,\kb}=2\langle \Phi^{+}_{2,\kb} \Phi^{-}_{0,-\kb}\rangle.
\end{equation}
This correlation imposes that all non-zero terms must have $\kb''=\kb$, which
removes the '$\ldots$' terms, so that (\ref{dd1}) becomes
\begin{equation}\label{dd1b}
\frac{F^{\Phi}_{2,0,\kb}}{\Delta \kb} =2\sum_{\kpb} M^2(\kb,\kpb)
 \frac{\langle \left|G_{\kb-\kpb}\right|^2\rangle}{\Delta \kb}
\frac{\langle\Phi^+_{0,\kb}\Phi^-_{0,-\kb}\rangle}{\Delta \kb } \langle f_2
\er^{\ir \sigma t}\rangle \Delta \kb,
\end{equation}
with
 \begin{equation}\label{lim3}
    \langle f_2 \er^{\ir \sigma t}\rangle
 =\frac{\upi  t}{8 \sigma
  \sigma'}\left\{\delta\left[\sigma'-(\sigma-\lb \bcdot \Ub)\right] + O(1)\right\}.
\end{equation}
Taking the limit when ${\Delta \kb}\rightarrow 0$, and neglecting $O(1)$ terms
yields
\begin{equation}\label{dE3b}
F^{\Phi}_{2,0}(t,\kb)=-\int_{\kpb}\frac{\upi t }{4 \sigma } M^2(\kb,\kpb)
  F^B(\kb-\kpb)\frac{F^{\Phi}_{0,0}(\kb)}{\sigma'}
\delta\left(\omega'-\omega\right) \dr  \kpb.
\end{equation}
Changing the spectral coordinates from $\kb'$ to $(\omega',\theta')$ allows a
simple removal of the singularity,
\begin{equation}\label{dE3c}
  %\frac{dE_{3,1}(\kb)}{dt}
F^{\Phi}_{2,0}(t,\kb)=-\int_{0}^{2\upi}\frac{\upi t }{4 \sigma } M^2(\kb,\kpb)
  F^B(\kb-\kpb)\frac{F^{\Phi}_{0,0}(\kb)}{\sigma'} \frac{k'}{Cg'+\kpb \bcdot \Ub/k'}
 \dr \theta'.
\end{equation}

 \section{Resonant wavenumber configuration for $U<<C_g$}
 Under the assumption $U<<C_g$, and for a current in the $x$ direction, the resonant conditions
 \begin{equation}
   \sigma'-\sigma = l_x U,\mbox{ and}
 \end{equation}
 yields the following Taylor expansion to first order in $\sigma'-\sigma$,
  \begin{equation}
  k'-k =(k_x'-k_x) \frac{U}{C_g} +O\left[k \left(\frac{U}{C_g}
  \right)^2\right].
 \end{equation}
We define,
 $r=k'$, $r_0=k$, $r\cos \theta=k'_x$, so that
 \begin{equation}
   r=r_0+\frac{U}{C_g}(r_0\cos \theta_0 -r\cos\theta),
 \end{equation}
and thus
 \begin{equation}\label{E2}
   r=\frac{P}{1+e\cos \theta}.
 \end{equation}
 This is the parametric equation of an ellipse of semi-major axis $a$, semi-minor axis $b$, half the foci distance
 $c$,
 and eccentricity $e$,
 with $  P=r_0+U/C_g r_0\cos \theta_0=b^2/a$, and
 $e=U/C_g=c/a$.
 The interaction between a surface wave with wavenumber
 $\kpb$ and a bottom component with wavenumber $\lb$
 excites a surface wave with the sum wavenumber
 $\kb=\kpb+\lb$. For a fixed $\kb$ and current $U$,
 in the limit of $U<<C_g$
 the resonant $\kpb$ and $\lb$ follow ellipses described by their polar equation (\ref{E2}),
 that reduce to circles for $U=0$.

\bibliographystyle{../references/jfm}
\bibliography{../references/wave}
\end{document}